\documentclass[aps,prb,twocolumn,epsf,showpacs,superscriptaddress]{revtex4-2}

\usepackage{graphicx}
\usepackage{dcolumn}
\usepackage{amsmath}
\usepackage{bm}
\usepackage{lipsum}
\usepackage{hyperref}

\DeclareUnicodeCharacter{2212}{-}

\begin{document}

\title{Quantum phase transition in overdoped La$_{2-x}$Sr$_{\mathbf{x}}$CuO$_{4}$
evinced by the superfluid density}

\author{T.~Schneider}\affiliation{Physik-Institut der Universit\"{a}t Z\"{u}rich,Winterthurerstrasse 190, CH-8057 Zürich, Switzerland}

\begin{abstract}
The superfluid density $\rho _{\Vert sf}\left( T\right) =\lambda _{\Vert
}^{-2}\left( T\right) $ of overdoped La$_{2-x}$Sr$_{\mathbf{x}}$CuO$_{4}$
thin films of high quality have been measured with $T_{c}$ (defined by the
onset of the Meissner effect ) from $5.1$ to $41.6$ K by Bo\v{z}ovi\'{c}
\textit{et al.} \cite{bosovic}. Given this $T_{c}$ the superfluid density $%
\rho _{\Vert sf}\left( T\right) $ shows no clear evidence of critical
fluctuations and no indication of vortex unbinding as $T\rightarrow T_{c}$.
Nevertheless, $\rho _{\Vert sf}\left( T\right) $ displays the behavior
expected for a quantum phase transition (QPT) in the $\left( 3+1\right) $%
D-xy dimensional universality class, with $\rho _{\Vert sf}\left( T\right)
\propto T_{c}^{-2}$ as $T\rightarrow 0$. However, this relation is also a
hallmark of dirty superconductors, treated in the mean-field approximation.
Here we attempt to clear out the nature of the suppression of $\rho _{\Vert
sf}\left( T\right) $ as $T_{c}\rightarrow 0$. Noting that for any finite
system the continuous transition will be rounded we perform a finite size
scaling analysis. It uncovers that the $\rho _{\Vert sf}\left( T\right) $
data are consistent with a finite length limited $3$D-xy transition. In some
films it is their thickness and in others their inhomogeneity that
determines the limiting length. Having established the precondition for the
occurrence of a QPT mapping on the $\left( 3+1\right) $D-xy model, we
explore the consistency with the hallmarks of this transition. In particular
with the relations $\rho _{\Vert sf}\left( T\right) /\rho _{\Vert s}\left(
0\right) =1-y_{c}T/T_{c}$, $T_{c}\propto 1/\lambda _{\Vert }\left( 0\right) $
as $T_{c}\rightarrow 0$ and $y_{c}=\alpha \lambda _{\parallel }^{2}\left(
0\right) T_{c}$, where $\alpha $ is the coefficient in $\rho _{\Vert
sf}\left( T\right) =\rho _{\Vert s}\left( 0\right) -\alpha T$. The emerging
agreement with these characteristics points clearly to a quantum
fluctuations induced suppression, revealing the crossover from the thermal
to the quantum critical regime as $T_{c}\rightarrow 0$. In the
classical-quantum mapping it corresponds to a $3$D to $(3+1)$D crossover.

\end{abstract}


\maketitle{}

\section{Introduction}
An essential unsolved puzzle in cuprate superconductivity is that, despite
the evidence of more conventional normal state properties \cite%
{proust,wang,hone2}, the superfluid density $\rho _{\Vert s}\left( T\right)
=\lambda _{\Vert }^{-2}\left( T\right) $ of overdoped La$_{2-x}$Sr$_{\mathbf{%
x}}$CuO$_{4}$ seems to be controlled by phase fluctuations \cite%
{bosovic,jaccard,rourke,mahmood}. Furthermore, the vanishing superfluid
density in overdoped La$_{2-x}$Sr$_{\mathbf{x}}$CuO$_{4}$ films is
consistent with $\rho _{\Vert s}\left( T\right) \propto $ $T_{c}^{2\text{ }}$
as $T_{c}\rightarrow 0$ \cite{bosovic}, expected for a quantum phase
transition in the $(3+1)$-dimensional (D)- xy universality class \cite%
{kim,ts}. This implies $3$D-xy critical behavior at any $T>0$, characteristic
for uncharged superfluids \cite{fisher}, as observed near optimum doping in
YBa$_{2}$Cu$_{3}$O$_{6+y}$ single crystals \cite{kamal,meingast} and other
cuprates \cite{ts,ariosa,tsk,tsg}. However, the measurements of $\rho _{\Vert
s}\left( T\right) $ on the overdoped La$_{2-x}$Sr$_{\mathbf{x}}$CuO$_{4}$
films show no clear evidence of 3D-xy critical behavior \cite{bosovic}.
Rather the approach to $T_{c}$ is mean-field-like. Taking this
mean-field-like behavior of $\rho _{\Vert s}\left( T\right) $ for granted
the suppression of $T_{c}$ is also attributable to pair breaking in a d-wave
superconductor. In the dirty limit it leads to the same scaling relation,
namely $\rho _{\Vert s}\left( T\right) \propto $ $T_{c}^{2}$ \cite{sun,kogan}.
The puzzle, namely what suppresses the superfluid density in overdoped
cuprates, is what we attempt to solve.

 First we perform a finite size scaling analysis of the superfluid density
data of Bo\v{z}ovi\'{c} \textit{et al.}\cite{bosovic}. It turns out that the
observed mean-field-like behavior is a finite size effect. It stems from the
fact that in an infinite and homogeneous 3D-xy system the transverse
correlation length diverges as $\xi _{\perp }^{t}\left( T\right) =\xi
_{\perp 0}^{t}t^{-\nu }$, where $t=\left( 1-T/T_{c}\right) $ and $\nu \simeq
2/3$ \cite{cucchieri}, while in the films $\xi _{\perp }^{t}\left(
T_{c}\right) $ is set by the limiting length $L_{c}$. In homogeneous films
it is determined by the effective film thickness $L_{cf}$ and in
inhomogeneous films by $L_{ci}<L_{cf}$, the size of the homogenous domains.
Nevertheless, in the films analyzed here, the $3$D-xy critical regime is
reached. This analysis provides estimates for the transition temperatures
and the critical amplitude $\lambda _{\parallel 0}\left( T_{c}\right) $ of
the infinite system appearing in $\lambda _{\parallel }^{-2}\left( T\right)
=\lambda _{\parallel 0}^{-2}t^{\nu }$ as $T\rightarrow T_{c}$. Because the
3D-xy transition temperatures are lower than those estimated from the
onset of the Meissner effect \cite{bosovic}, the mean-field-like
behavior of $\rho _{\Vert sf}\left( T\right) =\lambda _{\Vert f}^{-2}\left(
T\right) $ is traced back to a finite size rounded transition.
In addition the estimates for $T_{c}$ and $\lambda _{\parallel 0}$ uncover the
relationship $T_{c}\propto 1/\lambda _{\parallel 0}$ as $T_{c}\rightarrow 0$.
Accordingly, the system approaches the extreme type II limit $\ $($\lambda
_{\parallel }\rightarrow \infty $) as $T_{c}\rightarrow 0$. In this regime
it is expected that the static critical properties at zero external field
are dominated by phase fluctuations \cite{book,sudbo}.

Second, having established that 3D-xy fluctuations determine the finite
temperature critical behavior, limited by the finite size effect, we turn to
the analysis of the low temperature data extending down to $0.3$~K.
Therefore, an extrapolation is necessary to estimate $\lambda _{\parallel
}^{-2}\left( 0\right) $. It turns out that the linear extrapolation with $%
\lambda _{\parallel f}^{-2}\left( T\right) =\lambda _{\parallel }^{-2}\left(
0\right) -\alpha T$ fits the data up to $1$~K very well. Given then the
estimates of $T_{c}$, $\lambda _{\parallel }^{-2}\left( 0\right) ,$ $\alpha $
, and the $T_{c}$ dependence of the latter, their consistency with a QPT,
mapped on the $(3+1)$D-xy universality class, can now be tested. This
mapping requires that the scaling relations $T_{c}=f_{\Vert }(0)/\lambda
_{\parallel }\left( 0\right) $, $\alpha =y_{c}T_{c}/f_{\Vert }^{2}\left(
0\right) $ are satisfied and the data plotted as $\lambda _{\parallel
}^{2}\left( 0\right) /\lambda _{\parallel f}^{2}\left( T\right) $ \textit{vs}
. $T/T_{c}$ collapse on a single curve. In the limit $T/T_{c}\rightarrow 0$
and sufficiently small $T_{c}$ it should tend to the line $\lambda
_{\parallel }^{2}\left( 0\right) /\lambda _{\parallel f}^{2}\left( T\right)
=1-y_{c}T/T_{c}$ and in the $T/T_{c}\rightarrow 1$ limit to $t^{2/3}$.
However, the rounded transition leads around $T/T_{c}\approx 1$ to
characteristic deviations. $\lambda _{\parallel }^{2}\left( 0\right)
/\lambda _{\parallel f}^{2}\left( T\right) $ exceeds $T/T_{c}=1$ up to the
temperature set by the onset of the Meissner effect. As $T_{c}\rightarrow 0$
the rounding becomes stronger because the ratio $L_{c}/\xi _{\perp
}^{t}\left( T\right) =\left( L_{c}/\xi _{\perp 0}^{t}\right) t^{2/3}$,
controlling the finite size effect, decreases because $\xi _{\perp 0}^{t}$
grows like $\xi _{\perp 0}^{t}\propto 1/T_{c}$. On the other hand, as $%
T/T_{c}$ drops the collapse is no longer affected by the finite size effect.
The crossover from the thermal to the quantum fluctuation dominated regime
implies that $\lambda _{\parallel }^{-2}\left( T\right) =\lambda _{\parallel
0}^{-2}t^{2/3}$ approaches the low $T_{c}$ behavior and thus expands the
critical thermal regime $\lambda _{\parallel }^{-2}\left( T\right) =\lambda
_{\parallel }^{-2}\left( 0\right) t^{2/3}$ towards lower temperatures.
Noting that the QPT has been mapped on the (3+1)D-xy model it corresponds
here to a dimensional crossover from $3$D to $(3+1)$D. Our analysis reveals
that the $\rho _{\Vert sf}\left( T\right) =\lambda _{\Vert f}^{-2}\left(
T\right) $ data of the thin and overdoped La$_{2-x}$Sr$_{\mathbf{x}}$CuO$_{4}$
films are remarkably consistent with the outlined hallmarks of this
QPT. In summary, our analysis reveals that the suppression of the superfluid
density is driven by quantum phase fluctuations and uncovers the crossover from
the thermal to the quantum critical regime. The thermal regime shows finite
size limited 3D-xy criticality and the quantum regime is compatible with
$(3+1)$D-xy criticality.

 In Sec. II we sketch the theoretical background including the finite size
scaling approach, the scaling relations and scaling forms of the QPT mapped
on the classical $(3+1)$D-xy model. Sec. III is devoted to the analysis of
the superfluid density data of Bo\v{z}ovi\'{c} \textit{et al.}\cite{bosovic}.
We close with a brief summary and conclusions.

\section{Theoretical background}


A precondition for the occurrence of 3D-xy critical behavior in homogeneous
superconducting films with effective thickness $L_{c}$ originates from the
vortex interaction. Indeed, a Berezinskii \cite{berezinski}
-Kosterlitz-Thouless \cite{kosterlitz} (BKT) transition might occur, provided
that the Pearl length \cite{pearl} $\lambda _{pearl}=2\lambda _{\parallel
}^{2}\left( T_{BKT}\right) /L_{c}$ exceeds $L_{L}$, the lateral extent of
the film. In this case, a precondition for a BKT transition, the logarithmic
intervortex interaction, is satisfied. Invoking the Nelsom-Kosterlitz
relation \cite{nelson}, $T_{BKT}=\pi \Lambda L_{c}/\left( 2\lambda
_{\parallel }^{2}\left( T_{BKT}\right) \right) $ we obtain the precondition%
\begin{equation}
\lambda _{pearl}=\frac{\pi \Lambda }{T_{c}}>L_{L},  \label{eq1}
\end{equation}%
where $\Lambda =\Phi _{0}^{2}/\left( 16\pi ^{3}\text{k}_{B}\right) \simeq
6259~\mu m$~K. For the films considered here $L_{L}=1$cm and with that $%
T_{BKT}<\pi \Lambda /L_{L}\simeq 1.97$ K, which is considerably below $%
T_{c}\simeq 4.4$ K, the lowest value attained by Bo\v{z}ovi\'{c} \textit{et
al.}\cite{bosovic}.
In a homogeneous and sufficiently large system 3D-xy critical behavior
implies an infinite correlation length below $T_{c}$ \cite{hohenberg}. One
can extract an alternative diverging length the so called transverse
correlation length defined in terms of the helicity modulus $\Upsilon \left(
T\right) $ as $\xi ^{t}\left( T\right) =$ $\Upsilon \left( T\right) /k_{B}T$%
, where $\Upsilon \left( T\right) \lambda ^{2}\left( T\right) =\Phi
_{0}^{2}/\left( 16\pi ^{3}\right) $ and $\lambda $ denotes the penetration
depth \cite{kim,ts}. It probes the global superconducting phase coherence
across the system. In superconductors with uniaxial symmetry ($\lambda
_{x}=\lambda _{y}=\lambda _{\Vert }$, $\lambda _{z}=\lambda _{\perp }$) it
leads with%
\begin{equation}
\xi _{\perp }^{t}\left( t\right) =\xi _{\perp 0}t^{-\nu },\lambda _{\Vert
}^{2}\left( t\right) =\lambda _{\Vert 0}^{2}t^{-\nu },t=1-T/T_{c}\text{, }%
\nu \simeq 2/3\text{,}  \label{eq2}
\end{equation}%
to the universal relation \cite{ts}
\begin{equation}
T_{c}=\Lambda \xi _{\perp 0}^{t}/\lambda _{\parallel 0}^{2}.  \label{eq3}
\end{equation}%
It controls the $T_{c}$ dependence of the critical amplitudes $\xi _{\perp
0}^{t}$ and $\lambda _{\parallel 0}$.

 Other hallmarks of 3D-xy criticality follow from the derivative $d\lambda
_{\Vert }^{-2}/dT$ and the crossing of the curves $\lambda _{\Vert }\left(
T\right) ^{-2}$ and $\lambda _{\Vert }^{-3}\left( T\right) $. From Eq. (\ref%
{eq2}) we obtain
\begin{equation}
d\lambda _{\Vert }^{-2}/dT=-2/\left( 3\lambda _{\Vert 0}^{2}T_{c}\right)
t^{-1/3}.  \label{eq4}
\end{equation}%
Contrariwise there is no singularity in the mean-field treatment where $%
\lambda _{\Vert }^{2}\left( t\right) =\lambda _{\Vert 0}^{2}t^{-1}$ yields $%
d\lambda _{\Vert }^{-2}/dT=-1/\lambda _{\Vert 0}^{2}T_{c}$. Given 3D-xy
criticality (Eq. (\ref{eq2})) the curves $\lambda _{\Vert }\left( T\right)
^{-2}$ and $\lambda _{\Vert }^{-3}\left( T\right) $ cross at%
\begin{equation}
T_{cr}=T_{c}\left( 1-\left( \lambda _{\Vert 0}/a\right) ^{3}\right),
\label{eq5}
\end{equation}%
as long as $\lambda _{\Vert 0}/a<1$ with $a=1$ in units of $\lambda _{\Vert
0}$. In this case $\lambda _{\Vert }^{-3}\left( T\right) <\lambda _{\Vert
}\left( T\right) ^{-2}$ for $T<T_{cr}$ and $\lambda _{\Vert }^{-3}\left(
T\right) >\lambda _{\Vert }^{-2}\left( T\right) $ for $T>T_{cr}$.
Contrariwise, if $\lambda _{\Vert 0}/a>1$ there is no crossing point but
\begin{equation}
\lambda _{\Vert }\left( T\right) ^{-2}>\lambda _{\Vert }^{-3}\left( T\right).
\label{eq6}
\end{equation}%
Given then magnetic penetration depth data $\lambda _{\parallel }^{-2}\left(
T\right) $ of a homogeneous and infinite system the transition temperature
$T_{c}$ and the critical amplitude $\lambda _{\parallel 0}$ can be estimated
from fits to Eq. (\ref{eq2}), its derivative (Eq. (\ref{eq4})) and
crosschecked with the crossingpoint relations (\ref{eq5}) and (\ref{eq6}).

 In homogeneous films the outlined critical behavior will be modified because
the transverse correlation length $\xi _{\perp f}^{t}\left( T\right) $ does
not diverge at $T_{c}$ and is analytic for all $t$. However, for a
sufficiently large limiting length $L_{c}$ the finite size scaling theory
predicts a scaling behavior of the form \cite{barber,carac}
\begin{equation}
\xi _{\perp f}^{t}\left( t\right) /\xi _{\perp }^{t}\left( t\right) =g\left(
L_{c}/\xi _{\perp }^{t}\left( t\right) \right) .  \label{eq7}
\end{equation}%
In homogeneous films the limiting length is set by the effective film
thickness $L_{cf}$ $=L_{c}$ and in inhomogeneous films by $L_{c}=L_{ci}$ $%
<L_{cf}$, the size of the homogeneous domains. $g\left( x\right) $ is the
finite size scaling function. The critical $3$D-xy critical regime is
reached when $L_{c}/\xi _{\perp }^{t}\left( t\right) >1$ and $g\left(
L_{c}/\xi _{\perp }^{t}\left( t\right) \right) \rightarrow 1=\xi _{\perp
f}\left( t\right) /\xi _{\perp }\left( t\right) $. In contrast if $L_{c}/\xi
_{\perp }^{t}\left( t\right) <1$ the transition is rounded because $g\left(
L_{c}/\xi _{\perp }^{t}\left( t\right) \right) \rightarrow L_{c}/\xi _{\perp
}^{t}\left( t\right) $ and with that $\xi _{\perp f}\left( t\right)
\rightarrow L_{c}$. In principle, thick films are required for observing the
signatures of criticality such as Eq. (\ref{eq4}), and for accurate
determination of $T_{c}$, and the critical amplitudes $\xi _{\perp 0}$ and $%
\lambda _{\Vert 0}$. Noting that the divergence of $\xi _{\perp }^{t}\left(
t\right) $ is cut off at $\xi _{\perp }^{t}\left( t_{L_{c}}\right) =\xi
_{\perp 0}\left( 1-T_{L_{c}}/T_{c}\right) ^{-2/3}=L_{c}$ we obtain from $%
T_{L_{c}}/T_{c}=1-\left( \xi _{\perp 0}/L_{c}\right) ^{3/2}$ the precondition%
\begin{equation}
\xi _{\perp 0}<<L_{c},  \label{eq7b}
\end{equation}%
for a reliable finite size scaling analysis.
  Given the penetration depth data and the universal relation (\ref{eq3}) it
is convenient to transform the finite size scaling form (\ref{eq7}) to%
\begin{equation}
\lambda _{\parallel f}^{2}\left( t\right) /\lambda _{\parallel }^{2}\left(
t\right) =g\left( L_{\lambda \Vert }^{2}/\lambda _{\parallel }^{2}\left(
t\right) \right) ,  \label{eq8}
\end{equation}%
with%
\begin{equation}
L_{c}=L_{\lambda \Vert }^{2}T_{c}/\Lambda .  \label{eq9}
\end{equation}%
The critical $3$D-xy critical regime is reached when $L_{\lambda \Vert
}^{2}/\lambda _{\parallel }^{2}\left( t\right) <1$ and $g\left( L_{\lambda
\Vert }^{2}/\lambda _{\parallel }^{2}\left( t\right) \right) \rightarrow
1=\lambda _{\parallel f}^{2}\left( t\right) /\lambda _{\parallel }^{2}\left(
t\right) $. In contrast if $L_{\lambda \Vert }^{2}/\lambda _{\parallel
}^{2}\left( t\right) >1$ the transition is rounded because $g\left(
L_{\lambda \Vert }^{2}/\lambda _{\parallel }^{2}\left( t\right) \right)
\rightarrow L_{\lambda \Vert }^{2}/\lambda _{\parallel }^{2}\left( t\right)$
and $\lambda _{\parallel f}\left( t\right) \rightarrow L_{\lambda \Vert }$
by which $L_{c}$ can be determined with Eq. (\ref{eq9}). Using the estimates
for $T_{c}$, $1/\lambda _{\parallel 0}^{2}$ and $L_{\lambda \parallel }^{2}$
the plot $\left( \lambda _{\Vert f}\left( \left\vert t\right\vert \right)
/\lambda _{\Vert }\left( \left\vert t\right\vert \right) \right) ^{2}$
\textit{vs}. $L_{\lambda \Vert }^{2}/\lambda _{\Vert }^{2}\left( \left\vert
t\right\vert \right) $ should then let the experimental data\ ($\lambda
_{\Vert f}\left( \left\vert t\right\vert \right) $) collapse onto a single
line. For this to happen, the estimates for $T_{c}$, $1/\lambda
_{\parallel 0}^{2}$ and $L_{\lambda \parallel }^{2}=L_{c}\Lambda /T_{c}$
need to be correct. On the one hand, this plot enables the reliability of
the estimated values $T_{c}$, $\lambda _{\Vert 0}$ and $L_{c}$ to be
checked, and on the other hand it uncovers the nature of the limiting
length. If $L_{c}<L_{cf}$ it is traced back to the size of the homogenous
domains $L_{ci}$ and if $L_{c}\simeq L_{cf}$ to the effective film thickness
$L_{cf}$.

 In summary, the finite scaling analysis provides crosschecked
estimates of $T_{c}$, the critical amplitudes $\xi _{\bot 0}^{t}$ and $%
\lambda _{\Vert 0}$ of the infinite and homogeneous system and the $T_{c}$
dependence of the latter. Of particular interest is the resulting phase
transition line $T_{c}(1/\lambda _{\Vert 0})$ which shows how the finite
temperature superfluid density depends on $T_{c}$. Supposing that $T_{c}$
vanishes as
\begin{equation}
T_{c}=f_{\Vert 0}/\lambda _{\parallel 0},  \label{eq10}
\end{equation}%
the universal relation (\ref{eq3}) points to a QPT because the transverse correlation length
\begin{equation}
\xi _{\perp 0}^{t}=f_{\Vert 0}^{2}/\Lambda T_{c}.  \label{eq11}
\end{equation}%
diverges as $T_{c}\rightarrow 0$, while the temperature dependence of $%
\lambda _{\parallel }^{-2}\left( t\right) =\lambda _{\parallel
0}^{-2}t^{-2/3}$ and $\xi _{\perp }^{t}\left( t\right) $ is not affected.
In this case the system approaches the extreme Type II limit where
$\lambda_{\parallel }\rightarrow \infty $ as $T_{c}\rightarrow 0.$

Approaching the corresponding quantum critical point at $T=0$ we can
invoke the scaling theory of quantum critical phenomena \cite{kim,ts}.
A hallmark of a QPT mapped on the classical $(3+1)$ D-xy model is
\begin{equation}
T_{c}=f_{\Vert }\left( 0\right) /\lambda _{\parallel }\left( 0\right) ,
\label{eq12}
\end{equation}%
the quantum counterpart of Eq. (\ref{eq10}). Since D$=4$ is the upper
critical dimension of the XY model, mean-field critical behavior follows,
supplemented by logarithmic corrections \cite{sbsln,mssd4}. The scaling relations of this
transition include the quantum analog of Eq. (\ref{eq3})
\begin{equation}
T_{c}=R\Lambda \xi _{\perp }^{t}\left( 0\right) /\lambda _{\parallel
}^{2}\left( 0\right) ,  \label{eq13}
\end{equation}%
where $R$ is a nonuniversal coefficient. It yields with Eq. (\ref{eq12}) the
important relations
\begin{equation}
T_{c}=f_{\Vert }\left( 0\right) /\lambda _{\parallel }\left( 0\right)
=f_{\Vert }^{2}\left( 0\right) /\left( R\Lambda \xi _{\perp }^{t}\left(
0\right) \right) ,  \label{eq14}
\end{equation}%
In addition there is the scaling form \cite{kim,ts}
\begin{equation}
\lambda _{\parallel }^{2}\left( 0\right) /\lambda _{\parallel }^{2}\left(
T\right) =F\left( y_{c}T/T_{c}\right) ,\text{ }F\left( 0\right) =1
\label{eq15}
\end{equation}%
valid in the limit $T\rightarrow 0$ and low $T_{c}$. $y_{c}$ $=ab$ denotes
the universal critical value of the scaling variable $y=aT\delta ^{-1/2}$ at
which the scaling function $F\left( T=0,\delta \right) $ exhibits a
singularity at
\begin{equation}
T_{c}=b\delta ^{1/2}\propto 1/\lambda _{\parallel }\left( 0\right) ,
\label{eq15a}
\end{equation}%
where $\delta $ denotes the tuning parameter. A crossover to 3D-xy
criticality (Eq. (\ref{eq2})) then takes place as the temperature increases.
In $3$D it is the crossover from the quantum to the thermal critical regime
and in the quantum 3D- xy model mapped onto the ($3+1$)D-xy model a
dimensional crossover from ($3+1$) to $3$D. To check the compatibility with
scaling relations (\ref{eq12}), (\ref{eq13} and the scaling form (\ref{eq15})
estimates for $\lambda _{\parallel }\left( 0\right) $ are required. Given
the data extending down to $0.3$~K we have to relay on an extrapolation. A
linear temperature dependence $\lambda _{\parallel }^{-2}\left( T\right)
=\lambda _{\parallel }^{-2}\left( 0\right) -\alpha T$ leads with Eq. (\ref{eq15}) to the scaling form
\begin{eqnarray}
\lambda _{\parallel }^{2}\left( 0\right) /\lambda _{\parallel }^{2}\left(
T\right)  &=&F\left( y_{c}T/T_{c}\right) =1-y_{c}T/T_{c}  \nonumber \\
&=&1-y_{c}T_{c}\lambda _{\parallel }\left( 0\right) /f\left( 0\right) ,
\label{eq16}
\end{eqnarray}
and the scaling relation
\begin{equation}
y_{c}=\alpha \lambda _{\parallel }^{2}\left( 0\right) T_{c}=\alpha f_{\Vert
}\left( 0\right) \lambda _{\parallel }\left( 0\right) =\alpha f^{2}\left(
0\right) /T_{c},  \label{eq17}
\end{equation}%
setting a stringent constraint on the coefficient $\alpha $. Given then
estimates for $\alpha $ and $1/\lambda _{\parallel }^{2}\left( 0\right) $,
consistency with the $(3+1)$D- xy QPT requires that the
data, plotted as $\lambda _{\parallel }^{2}\left( 0\right) /\lambda
_{\parallel f}^{2}\left( T\right) $ \textit{vs}. $T\lambda _{\parallel
}\left( 0\right) $ or \textit{vs}. $T/T_{c}$, collapse for sufficiently low
$T_{c}$ as $T\rightarrow 0$ on the line given by Eq. (\ref{eq16}), provided
that the estimates for $\lambda _{\parallel }\left( 0\right) $, $\alpha $
and $T_{c}$ are correct and $\alpha $ scales according to Eq. (\ref{eq17}) as
\begin{equation}
\alpha =y_{c}T_{c}/f_{\Vert }^{2}\left( 0\right) .  \label{eq18}
\end{equation}
This linear temperature dependence $\left. d\lambda _{\Vert
}/dT\right\vert _{T\rightarrow 0}=\alpha \lambda _{\Vert }^{3}\left(
0\right) /2$ was proposed by several authors in terms of phase fluctuations
\cite{stroud,coffey,deutscher}.

 Finally, the relationship between the zero temperature penetration depth
 $\lambda _{\parallel }\left( 0\right) $, the correlation length $\xi _{\perp
}\left( 0\right) $ and the respective critical amplitudes, $\lambda
_{\parallel 0}$ and $\xi _{\perp 0}^{t}$, need to be clarified. The
crossover from quantum to classical critical behavior implies $\lambda
_{\parallel }^{2}\left( 0\right) /\lambda _{\parallel }^{2}\left( T\right)
=F\left( y_{c}T/T_{c}\right) \rightarrow $ $\lambda _{\parallel
0}^{2}/\lambda _{\parallel }^{2}\left( T\right) t^{2/3}$ as $T/T_{c}$
increases and with that the crossover from $\lambda _{\parallel }\left(
0\right) $ to the finite temperature amplitude $\lambda _{\parallel 0}$.
This crossover does not modify the characteristic temperature dependence of
3D-xy criticality ($t^{2/3}$) but reflects the quantum fluctuation induced
change of the critical amplitude $\lambda _{\parallel 0}$. With
$\lambda_{\parallel 0}\rightarrow \lambda _{\parallel }\left( 0\right)$ as
$T_{c}\rightarrow 0$ the scaling relations (\ref{eq10}) and (\ref{eq12})
reduce then to
\begin{equation}
T_{c}=f_{\Vert }\left( 0\right) /\lambda _{\parallel }\left( 0\right)
{}=f_{\Vert }\left( 0\right) /\lambda _{\parallel 0}.  \label{eq19}
\end{equation}%
Combining this result with the scaling relations (\ref{eq3}), (\ref{eq14})
we obtain the desired relationship
\begin{eqnarray}
T_{c} &=&f_{\Vert }\left( 0\right) /\lambda _{\parallel }\left( 0\right)
=f_{\Vert }^{2}\left( 0\right) /\left( R\Lambda \xi _{\perp }^{t}\left(
0\right) \right)  \nonumber \\
&=&f_{\Vert }\left( 0\right) /\lambda _{\parallel 0}=f_{\Vert }^{2}\left(
0\right) /\Lambda \xi _{\perp 0}^{t}.  \label{eq20}
\end{eqnarray}%

An essential implication of this crossover is the growth of the amplitude
$\xi _{\perp 0}^{t}$ as $T_{c}\rightarrow 0$. It crosses over to
$\xi _{\perp}^{t}\left( t\right) =\xi _{\perp 0}^{t}t^{-2/3}=\left( f_{\Vert }^{2}\left(
0\right) /\Lambda T_{c}\right) t^{-2/3}$ which leads to an inevitable
enhancement of the finite size effect controlled by the ratio $L_{c}/\xi
_{\perp }^{t}\left( T\right) =\left( L_{c}/\xi _{\perp 0}^{t}\right) t^{2/3}$
(Eq. (\ref{eq7})) because $L_{c}/\xi _{\perp }^{t}\left( T\right) =\left(
L_{c}/\xi _{\perp 0}^{t}\right) t^{2/3}\propto L_{c}T_{c}t^{2/3}$ as
$T_{c}\rightarrow 0$. In contrast, the zero temperature counterpart
$L_{c}/\xi _{\perp }^{t}\left( 0\right) $ remains large in spite of
$\xi_{\perp }^{t}\left( 0\right) \propto 1/T_{c}$ for any reasonable $L_{c}$
because $t^{2/3}$ does not enter.

Another manifestation of the quantum fluctuations comes from their effect on
the condensate and Cooper-pair density. In the Ginzburg-landau version of
the D-xy-model, where amplitude $\left\vert \Psi \left( R\right) \right\vert
$and phase $\varphi \left( R\right) $ are taken into account, these
densities can be expressed in terms of the complex order parameter $\Psi
\left( R\right) \exp (i\varphi \left( R\right) $.\ The superfluid density $%
\rho _{\Vert s}\left( 0\right) =\lambda _{\Vert }^{-2}\left( 0\right) $, the
helicity modulus $\Upsilon _{\Vert }$, the condensate density $\left\langle
\left\vert \Psi \right\vert \right\rangle ^{2}$ and the Cooper-pair
counterpart $\left\langle \left\vert \Psi \right\vert ^{2}\right\rangle $
are found to scale as \cite{sudbo}%
\begin{eqnarray}
\rho _{\Vert s}\left( t\right)  &=&\lambda _{\Vert 0}^{-2}t^{\nu }=(16\pi
^{3}/\Phi _{0}^{2})\Upsilon _{\Vert 0}t^{\nu }\neq \left( \hbar
^{2}/2m_{\Vert }\right) \left\vert \left\langle \Psi \right\rangle
\right\vert ^{2},  \nonumber \\
\text{ }\left\langle \left\vert \Psi \right\vert \right\rangle ^{2} &\propto
&t^{2\beta },\text{ }\left\langle \left\vert \Psi \right\vert
^{2}\right\rangle >\left\langle \left\vert \Psi \right\vert \right\rangle
^{2},\text{ }\left\langle \left\vert \delta \Psi \right\vert
^{2}\right\rangle >0.  \label{eq20a}
\end{eqnarray}%
Accordingly, the superfluid and condensate densities vanish as $T\rightarrow
T_{c}$ while the Cooper pair density remains finite at and above $T_{c}$.
Indeed, at $T_{c}$ it scales as $\left\langle \left\vert \delta \Psi
\right\vert ^{2}\right\rangle _{T=T_{c}}\propto T_{c}$. The difference between
the superfluid and condensate density stems from the fact
that in the 3D-xy model $\nu =2\beta -\eta $ \cite{mefisher,josephson} with
$\eta \simeq 0.0381$ \cite{hasenbusch}. The appearance of 3D-xy critical
behavior stems from the fact that in extreme type II superconductors the
amplitude fluctuations are dominated by phase fluctuations over a sizeable
temperature regime \cite{sudbo}.  Note that in the QPT considered here, where
$T_{c}\propto 1/\lambda _{\parallel }\left( 0\right) {}\propto 1/\lambda
_{\parallel 0}$ (Eq. (\ref{eq19})), the extreme type II limit ($\lambda
_{\parallel }\rightarrow \infty $) is attained as $T_{c}\rightarrow 0$.

The mapping of the QPT on the $\left( 3+1\right) $ D-xy
model implies mean-field critical behavior. Therefore, taking the quantum
fluctuations in the 3D-xy model into account, the superfluid density $\rho
_{\Vert s}\left( 0\right) =\lambda _{\Vert }^{-2}\left( 0\right) $, the
helicity modulus $\Upsilon _{\Vert }\left( 0\right) $, the condensate
density $\left\langle \left\vert \Psi \right\vert \right\rangle _{T=0}^{2}$
and the Cooper-pair counterpart $\left\langle \left\vert \Psi \right\vert
^{2}\right\rangle _{T=0}$ scale as
\begin{eqnarray}
\rho _{\Vert s}\left( 0\right)  &=&\lambda _{\Vert }^{-2}\left( 0\right)
=(16\pi ^{3}/\Phi _{0}^{2})\Upsilon _{\Vert }\left( 0\right)   \nonumber \\
&=&\left( \hbar ^{2}/2m_{\Vert }\right) \left\vert \left\langle \Psi
\right\rangle \right\vert _{T=0}^{2},  \nonumber \\
\left\langle \left\vert \Psi \right\vert \right\rangle _{T=0}^{2}
&=&\left\langle \left\vert \Psi \right\vert ^{2}\right\rangle _{T=0}\propto
\delta \propto T_{c}^{2\text{ }},  \label{eq20b}
\end{eqnarray}
where we used Eq. (\ref{eq15a}). $\delta $ denotes the tuning parameter of
the QPT. Remarkably, not only the superfluid density is suppressed by
quantum fluctuations as $T_{c}\rightarrow 0$ but also the condensation and
Cooper pair density.

\section{Data Analysis}\label{seq:Analysis}

\begin{figure}[t]
\includegraphics[width = 1\columnwidth]{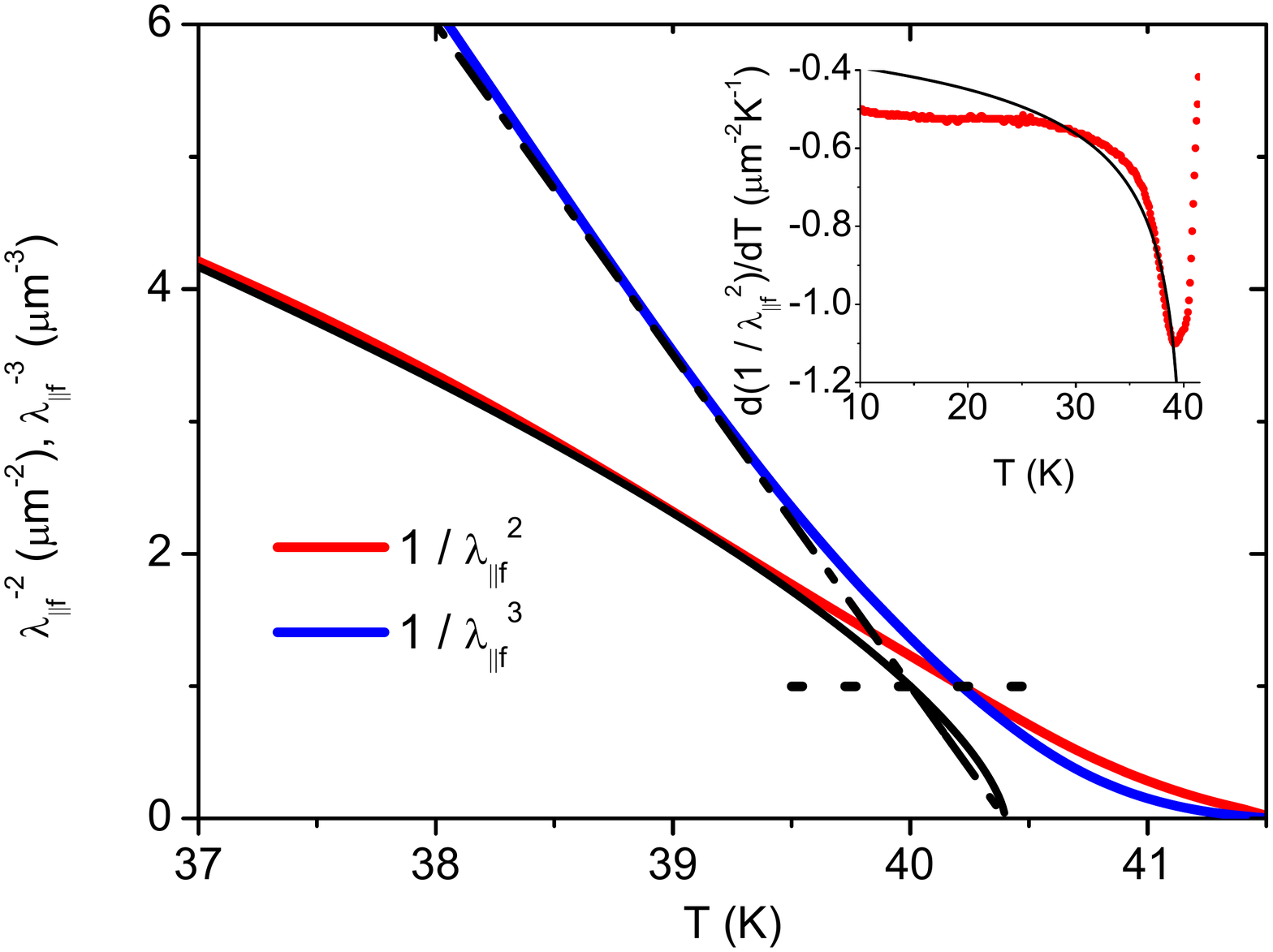}
\caption{$\lambda _{\parallel f}^{-2}\left( T\right) $, $\lambda _{\parallel
f}^{-3}\left( T\right) $ \textit{vs}. $T$ \ of film S1 marked by the red and
blue lines. The solid black and the dash-dot lines mark the 3D-xy critical
behavior \ in terms of $\lambda _{\parallel }^{-2}\left( T\right) =\lambda
_{\parallel 0}^{-2}t^{-2/3}$, $\lambda _{\parallel }^{-3}\left( T\right)
=\lambda _{\parallel 0}^{-3}t$ with $\lambda _{\parallel 0}^{-2}$ and $T_{c}$
values listed in Table I. The crossing point of the experimental data occurs
at $T_{crf}\simeq 40.2$~K and the 3D-xy counterpart at $T_{cr}\simeq 40$~K .
The inset shows $d\lambda _{\parallel f}^{-2}\left( T\right) /dT$ vs. $T$
and the solid line marks the 3D-xy critical behavior (Eq. (\ref{eq4})) with
the parameters listed in Table \ref{table1}.}
\label{fig1}
\end{figure}

We are now prepared to analyze the penetration depth data. In the nearly
optimally doped film S1 evidence for the relevance of 3D-xy fluctuations
emerges from Fig. \ref{fig1}, revealing a crossing point in the plots $\lambda
_{\parallel f}^{-2}\left( T\right) $, $\lambda _{\parallel f}^{-3}\left(
T\right) $ \textit{vs}. $T$ (Eq. (\ref{eq6})). Indeed, $\lambda _{\Vert
}^{-2}\left( t\right) =\lambda _{\Vert 0}^{-2}\left( 1-T/T_{c}\right) ^{2/3}$
is a characteristic of 3D-xy criticality. The experimental data is seen to
cross at $T_{crF}\simeq 40.2$~K, a little higher than $T_{cr}\simeq 40$~K,
the value that corresponds to the estimated $3$D-xy critical behavior $%
T_{c}=40.4$~K (Eq. (\ref{eq5}) Table \ref{table1}). This and the fact that $\lambda
_{\parallel f}^{-2}\left( T\right) $ does not vanish at and extends beyond $%
T_{c}$ points to a finite size effect that rounds the sharp
3D-xy-transition, marked by the black line. It implies that the 3D-xy
transition temperature $T_{c}=40.4$~K is considerably lower than the $%
T_{c}\simeq $ $41.6$~K, defined by the onset of the Meissner effect\cite%
{bosovic}. Additional evidence for a finite size induced rounding of the
transition comes from $d\lambda _{\parallel f}^{-2}\left( T\right) /dT$
\textit{vs}. $T$ , shown in the inset of Fig. \ref{fig1}. As the estimated
$T_{c}=40.4 $~K is approached from below the data tends to the
characteristic 3D-xy singularity, $d\lambda _{\Vert }^{-2}/dT=-2/\left(
3\lambda _{\Vert 0}^{2}T_{c}\right) t^{-1/3}$ (Eq. (\ref{eq4})), marked by
the solid line. However, close to $T_{c}$ the singularity is cut off and
$\lambda _{\parallel f}$ approaches $\lambda _{\parallel f}\left(
T_{c}\right) =$ $L_{\lambda \Vert }\propto L_{c}$(Eqs. (\ref{eq8}), (\ref{eq9})).

\begin{figure}[t]
\includegraphics[width = 1\columnwidth]{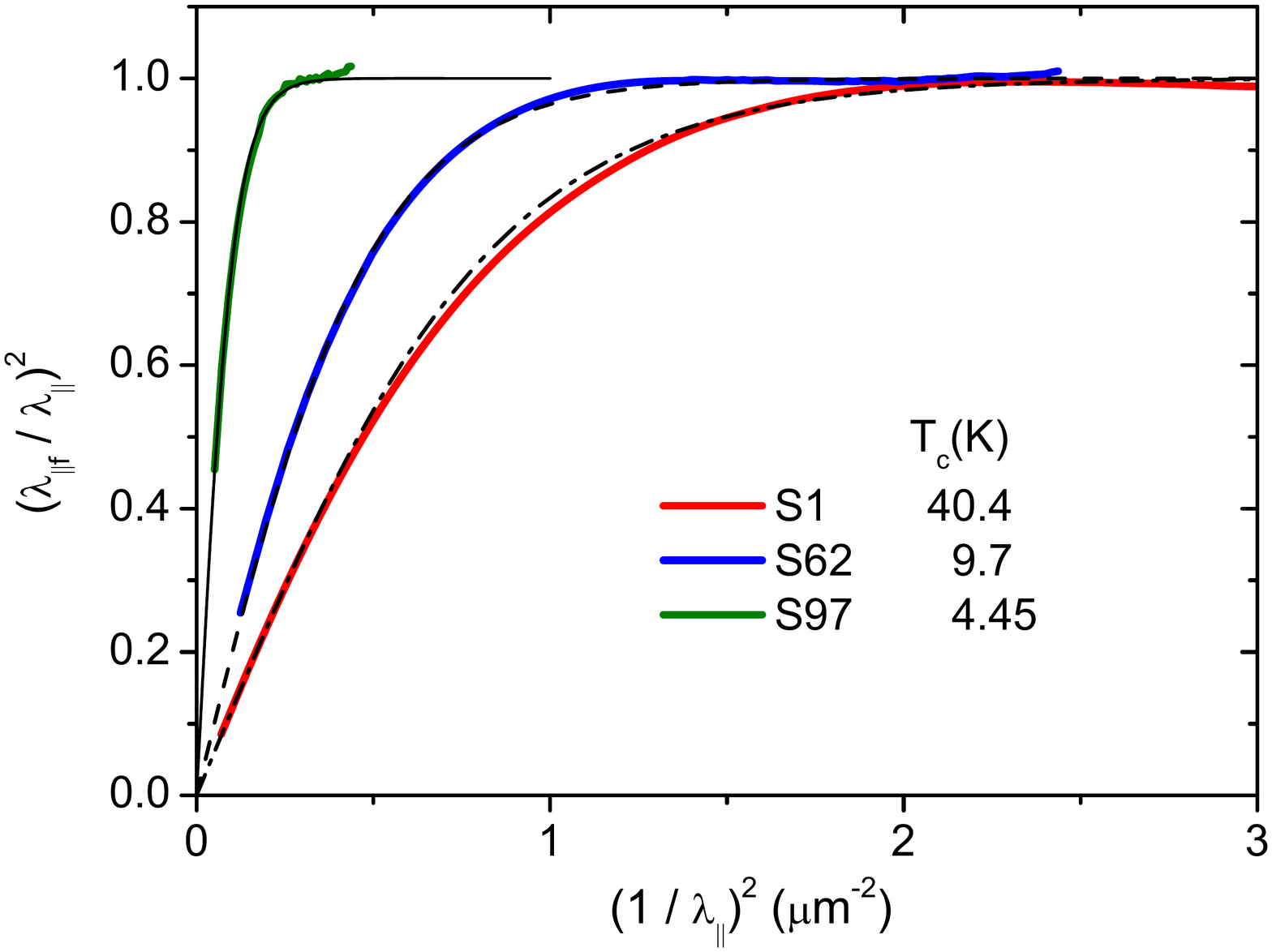}
\caption{Finite size scaling plots $\lambda _{\parallel f}^{2}\left( t\right)
/\lambda _{\parallel }^{2}\left( t\right) $ \textit{vs}. $1/\lambda
_{\parallel }^{2}\left( t\right) $ for the films S1, S62, and S97. The
limiting behavior $\left( \lambda _{\parallel f}\left( t\right) /\lambda
_{\parallel }\left( t\right) \right) ^{2}\rightarrow \left( L_{\lambda \Vert
}/\lambda _{\parallel }\left( t\right) \right) ^{2}$ yields $L_{\lambda
\Vert }^{2}$ and with Eq. (\ref{eq13}) the estimates for the limiting length
$L_{c}$ listed in Table \ref{table1}. The black solid, dashed, and
dashed-dot lines are $\left( \lambda _{\parallel f}\left( t\right) /\lambda
_{\parallel }\left( t\right) \right) ^{2}=\tanh (L_{\lambda \Vert
}^{2}/\lambda _{\parallel }^{2}\left( t\right) )$ which describe the
crossover from the finite size $\left( \lambda _{\parallel f}\left( t\right)
/\lambda _{\parallel }\left( t\right) \right) ^{2}<1$ to the 3D-xy critical
regime $\left( \lambda _{\parallel f}\left( t\right) /\lambda _{\parallel
}\left( t\right) \right) ^{2}\rightarrow 1$ remarkably well.}
\label{fig2}
\end{figure}

To unravel the nature of the weakly rounded transition we performed the
finite size scaling analysis of the data. Fig. \ref{fig2} shows $\lambda _{\parallel
f}^{2}\left( t\right) /\lambda _{\parallel }^{2}\left( t\right) $ \textit{vs}
. $1/\lambda _{\parallel }^{2}\left( t\right) $ for the films S1, S62, and
S97. From the limiting behavior $\lambda _{\parallel f}^{2}\left( t\right)
/\lambda _{\parallel }^{2}\left( t\right) \rightarrow L_{\lambda \Vert
}^{2}/\lambda _{\parallel }^{2}\left( t\right) $ (Eq. (\ref{eq8})) we derive
$L_{\lambda \Vert }^{2}\simeq 1.2$ $\mu m^{2}$ for the film S1, yielding
with Eq.(\ref{eq13}) $L_{c}\simeq 8$~nm for the limiting length, which is
slightly larger than the quoted film thickness $L_{cf}=6.6$~nm \cite{bosovic}.
Nevertheless, the $3$D-xy transition is only slightly rounded in this thin film
because the amplitude of the transverse correlation
length $\xi _{\parallel 0}=0.3$~nm (Table \ref{table1}) is small compared to
the film thickness (Eq. (\ref{eq7b})). Therefore, the analysis compiled in
Figs. \ref{fig1} and \ref{fig2} shows clearly that this film is homogeneous
and  exhibits due to its thickness a weakly rounded 3D-xy-transition.
Moreover, the approximately linear temperature dependence of $1/\lambda
_{\parallel f}^{2}$, over a rather wide temperature range, is then traced
back to the film thickness induced finite size effect and is not the
fingerprint of a d-wave superconductor in the dirty limit \cite%
{hone2,sun,kogan}. Nevertheless, associating the drop of $T_{c}$ with
increased dopant concentration, it seems plausible that in the limit
$T_{c}\rightarrow 0$ the size of the homogenous domains instead of the film
thickness set the limiting length.

\begin{table}[htb]
 \caption[~]{Estimates for $T_{c}$, the 3D-xy critical amplitudes $\lambda
 _{\parallel 0}$, $\xi _{\bot 0}^{t}$ (Eq. (\ref{eq2})), the limiting length $%
 L_{\lambda \Vert }$ of the penetration depth $\lambda _{\parallel }\left(
 T\right) $, the resulting limiting length $L_{c}$ (Eq. (\ref{eq9})), the
 zero temperature penetration depth $\lambda _{\parallel 0}\left( 0\right) $,
 and the coefficient $\alpha $ entering the linear temperature dependence of $%
 \lambda _{\parallel }^{-2}\left( T\right) $ in the limit $T\rightarrow 0$
 (Eq. (\ref{eq16})). The quoted film thickness is $L_{cf}=6.6$~nm \cite{bosovic}.}

 \label{table1}
\begin{tabular}{|c|l|l|l|l|l|l|l|}
\hline
& $T_{c}$ & $\lambda _{\parallel 0}^{-2}$ & $\xi _{\bot 0}^{t}$ & $%
L_{\lambda \Vert }^{2}$ & $L_{c}$ & $\lambda _{\parallel
}^{2}\left( 0\right) $ & $\alpha $ \\ \hline
& K & $\mu $m$^{-2}$ & nm & $\mu $m$^{2}$ & nm & $\mu $m$^{-2}$ & $\mu $m$%
^{-2}$~K$^{-1}$ \\ \hline
\multicolumn{1}{|l|}{S1} & 40.4 & 21.70 & 0.30 & 1.2 & 7.7 & $\simeq $23.20
& $\simeq $0.35 \\ \hline
\multicolumn{1}{|l|}{S62} & 9.7 & 2.62 & 0.59 & 2.0 & 3.1 & 2.54 & 0.124 \\
\hline
\multicolumn{1}{|l|}{S73} & 6.73 & 1.10 & 0.98 & 4.0 & 4.3 & 1.07 & 0.080 \\
\hline
\multicolumn{1}{|l|}{S87} & 5.64 & 0.76 & 1.19 & 8.5 & 7.7 & 0.75 & 0.082 \\
\hline
\multicolumn{1}{|l|}{S97} & 4.45 & 0.49 & 1.45 & 9.5 & 6.8 & 0.47 & 0.061 \\
\hline
\multicolumn{1}{|l|}{S98} & 4.27 & 0.47 & 1.45 & 10.0 & 6.8 & 0.46 & 0.070
\\ \hline
\end{tabular}
\end{table}

\begin{figure}[t]
\includegraphics[width = 1 \columnwidth]{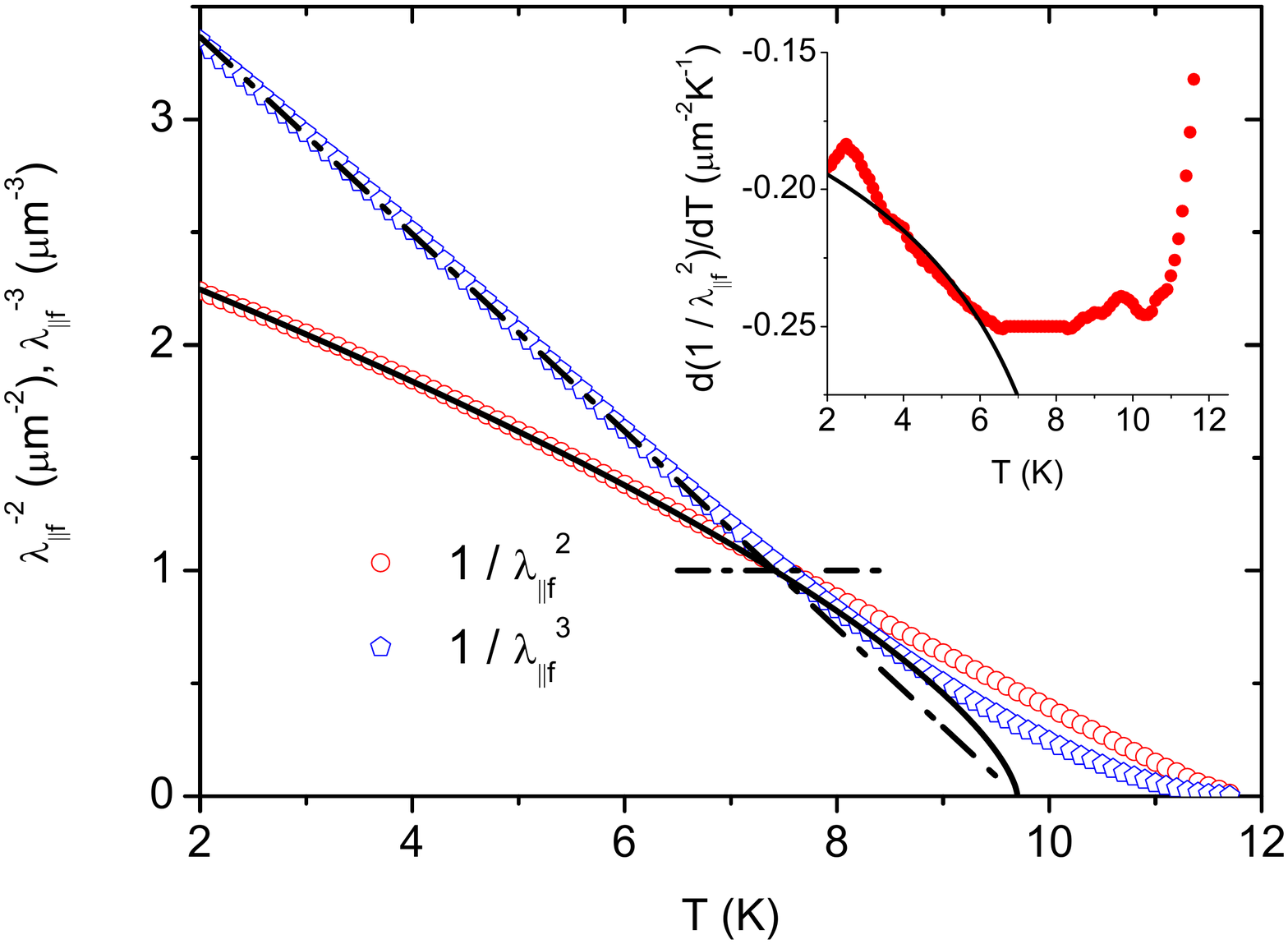}
\caption{$\lambda _{\parallel f}^{-2}\left( T\right) $, $\lambda _{\parallel
f}^{-3}\left( T\right) $ \textit{vs}. $T$ \ of film S62 marked by the red
and blue lines. The solid black and the dash-dot lines mark the 3D-xy
critical behavior \ in terms of $\lambda _{\parallel }^{-2}\left( T\right)
=\lambda _{\parallel 0}^{-2}t^{-2/3}$, $\lambda _{\parallel }^{-3}\left(
T\right) =\lambda _{\parallel 0}^{-3}t$ with $\lambda _{\parallel 0}^{-2}$
and $T_{c}$ values listed in Table \ref{table1}. The crossing point of the
experimental data occurs at $T_{crf}\simeq 7.5$~K and the 3D-xy counterpart
at $T_{cr}\simeq 7.4$~K .}
\label{fig3}
\end{figure}

To clarify this issue we turn to the analysis of the films S62 and S97 with
$T_{c}\simeq 9.7$~K and $T_{c}\simeq 4.45$~K, respectively. Fig. \ref{fig3}
shows $\lambda _{\parallel f}^{-2}\left( T\right) $, $\lambda _{\parallel
f}^{-3}\left( T\right) $ \textit{vs}. $T$ \ of the film S62. The crossing
point at $T_{crf}\simeq 7.5$~K uncovers again the presence of 3D-xy
fluctuations. However, the deviations from the estimated 3D-xy critical
behavior near and above $T_{c}$ reveal a strongly rounded transition. This
behavior is also visible in $d\lambda _{\Vert }^{-2}\left( T\right) /dT$
shown in the inset of Fig. \ref{fig3}. The critical behavior $d\lambda
_{\Vert }^{-2}\left( T\right) /dT=-2/\left( 3\lambda _{\Vert
0}^{2}T_{c}\right) t^{-1/3}$ (Eq. (\ref{eq4})) is attained at rather low
temperatures. Both phenomena are consistent with the crossover from the
thermal to the quantum critical regime where the transverse correlation
length scales as $\xi _{\perp }^{t}\left( t\right) \propto t^{-2/3}/T_{c}$
(Eq.(\ref{eq20})). Indeed, the crossover $\xi _{\perp }^{t}\left( t\right)
=\xi _{\perp 0}^{t}t^{-2/3}\rightarrow \xi _{\perp }^{t}\left( 0\right)
t^{-2/3}$ extends the 3Dxy critical regime to lower temperatures and
enhances the finite size effect for fixed $L_{c}$, because $L_{c}/\xi
_{\perp }^{t}\left( t\right) =L_{c}t^{2/3}/\xi _{\perp 0}^{t}\rightarrow
L_{c}t^{2/3}/\xi _{\perp }^{t}\left( 0\right) \propto L_{c}T_{c}t^{2/3}$ and
weak rounding requires that $L_{c}/\xi _{\perp }^{t}\left( t\right) >1$(Eq. (\ref{eq7})).
On the other hand, it seems plausible that stronger rounding
with reduced $T_{c}$ is attributable to a shrinkage of the homogenous
domains with increasing doping. In fact the limiting length $L_{c}\simeq 3.1$~nm
(Table \ref{table1}), derived from the finite size scaling plot shown in
Fig. \ref{fig2}, is considerably smaller than the quoted film thickness
$L_{cf}=6.6$~nm \cite{bosovic} and appears to confirm this scenario.
Nevertheless, Fig. \ref{fig3} shows that the film attains the
3D-xy critical regime. Supposing that inhomogeneities stem from
superconducting and metallic regions, with the fraction of metallic regions
increasing on overdoping \cite{wang}, the limiting length $L_{c}$ is expected
to decrease as $T_{c}\rightarrow 0$. To clarify this issue we turn to the
analysis of the film S97 with $T_{c}\simeq 4.45$~K.

\begin{figure}[t]
\includegraphics[width = 1 \columnwidth]{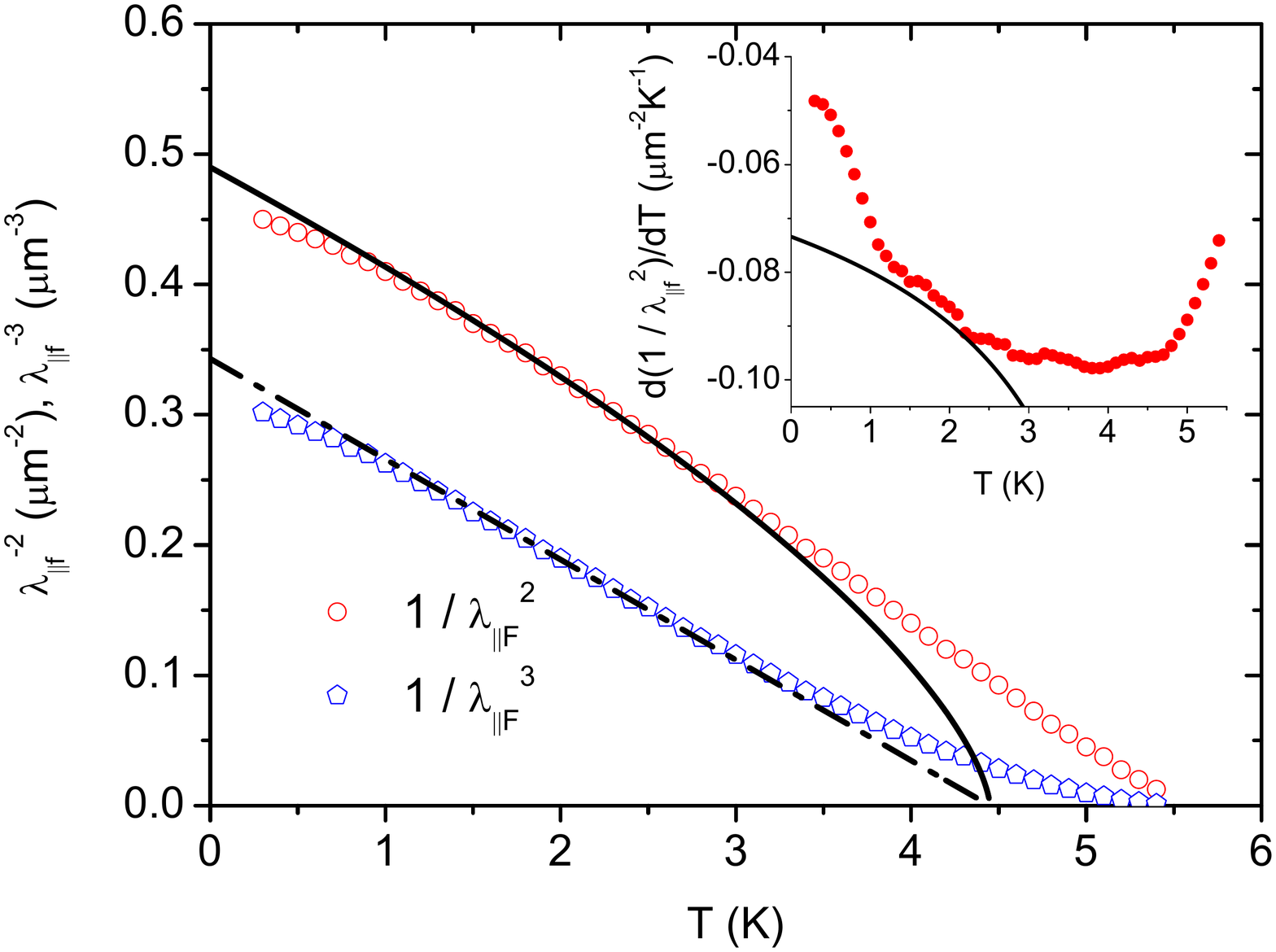}
\caption{$\lambda _{\parallel f}^{-2}\left( T\right) $, $\lambda _{\parallel
f}^{-3}\left( T\right) $ vs. $T$ \ of film S97 marked by the red and blue
lines,. The solid black line and the dash-dot line indicate the estimated
3D-xy critical behavior, $\lambda _{\parallel }^{-2}\left( T\right) =\lambda
_{\parallel 0}^{-2}t^{-2/3}$, $\lambda _{\parallel }^{-3}\left( T\right)
=\lambda _{\parallel 0}^{-3}t$ with $\lambda _{\parallel 0}^{-2}$ and $T_{c}$
listed in Table \ref{table1}. There is no crossing point because $\lambda
_{\parallel 0}^{-2}\simeq 0.49~\mu $m$^{-2}<1\mu $m$^{-2}$. Nevertheless
there is still evidence for 3D-xy critical behavior emerging from $d\lambda
_{\parallel F}^{-2}\left( T\right) /dT$ vs. $T$ shown in the inset. The
solid line represents the corresponding 3D-xy critical behavior (Eq. (\ref%
{eq4})) with the parameters listed in Table \ref{table1}.}
\label{fig4}
\end{figure}

Fig. \ref{fig4} shows $\lambda _{\parallel f}^{-2}\left( T\right) $, $%
\lambda _{\parallel f}^{-3}\left( T\right) $ vs. $T$ for the film S97. The
solid black line and the dash-dot line indicate the estimated 3D-xy critical
behavior of , $\lambda _{\parallel }^{-2}\left( T\right) $ and $\lambda
_{\parallel }^{-3}\left( T\right) $. Essential features include the absence
of the crossing point, the strongly rounded transition and the associated
shift of the attainable 3D-xy critical regime to lower temperatures. The
absence of the crossing point is according to Eq. (\ref{eq5}) a consequence
of 3D-xy criticality because the critical amplitude $\lambda _{\parallel
0}^{-}$ of this film satisfies the condition $\lambda _{\parallel
0}^{-2}\simeq 0.49~\mu $m$^{-2}<1~\mu $m$^{-2}$ whereupon $\lambda _{\Vert
}\left( T\right) ^{-2}$ $>\lambda _{\Vert }^{-3}\left( T\right) $ follows
(Eq. (\ref{eq6})). To clarify the origin of the strongly rounded transition,
also visible in $d\lambda _{\parallel f}^{-2}\left( T\right) /dT$ vs. $T$
depicted in the inset of Fig.\ \ref{fig4}, we turn to the results of the
finite size scaling analysis shown in Fig. \ref{fig2}. The essential result
is that the finite size limited 3D-xy criticality mimics the heavily smeared
transition well. But in this case with a limiting length $L_{c}=7.1$~nm
(Table \ref{table1}), compatible with the quoted film thickness. Therefore,
in spite of the pronounced overdoping, the finite size scaling analysis
uncovers the homogeneity of this film. We have carried out the same analysis
for the films S73, S87 and S98. The result is that all show agreement with a
film thickness limited 3D-xy criticality. Comparing the limiting lengths $%
L_{c}$ listed in Table \ref{table1} with the film thickness, $L_{cf}=6.6$~nm
it follows that the limiting length responsible for the rounded 3D-xy
transitions in S87 and S98 stems from the film thickness, because
$L_{c}\simeq L_{cF}$, while in S73, where $L_{c}$ is considerably smaller
than $L_{cf}$, inhomogeneities set the limiting length. However, the missing
systematic correlation between $T_{c}$ and $L_{c}$ suggests that
$L_{c}<L_{cf} $ simply points to worse film quality and not to an unavoidable
doping-induced phase separation, according to which the size of the
homogeneous domains should shrink with reduced $T_{c}$.

\begin{figure}[t]
\includegraphics[width = 1 \columnwidth]{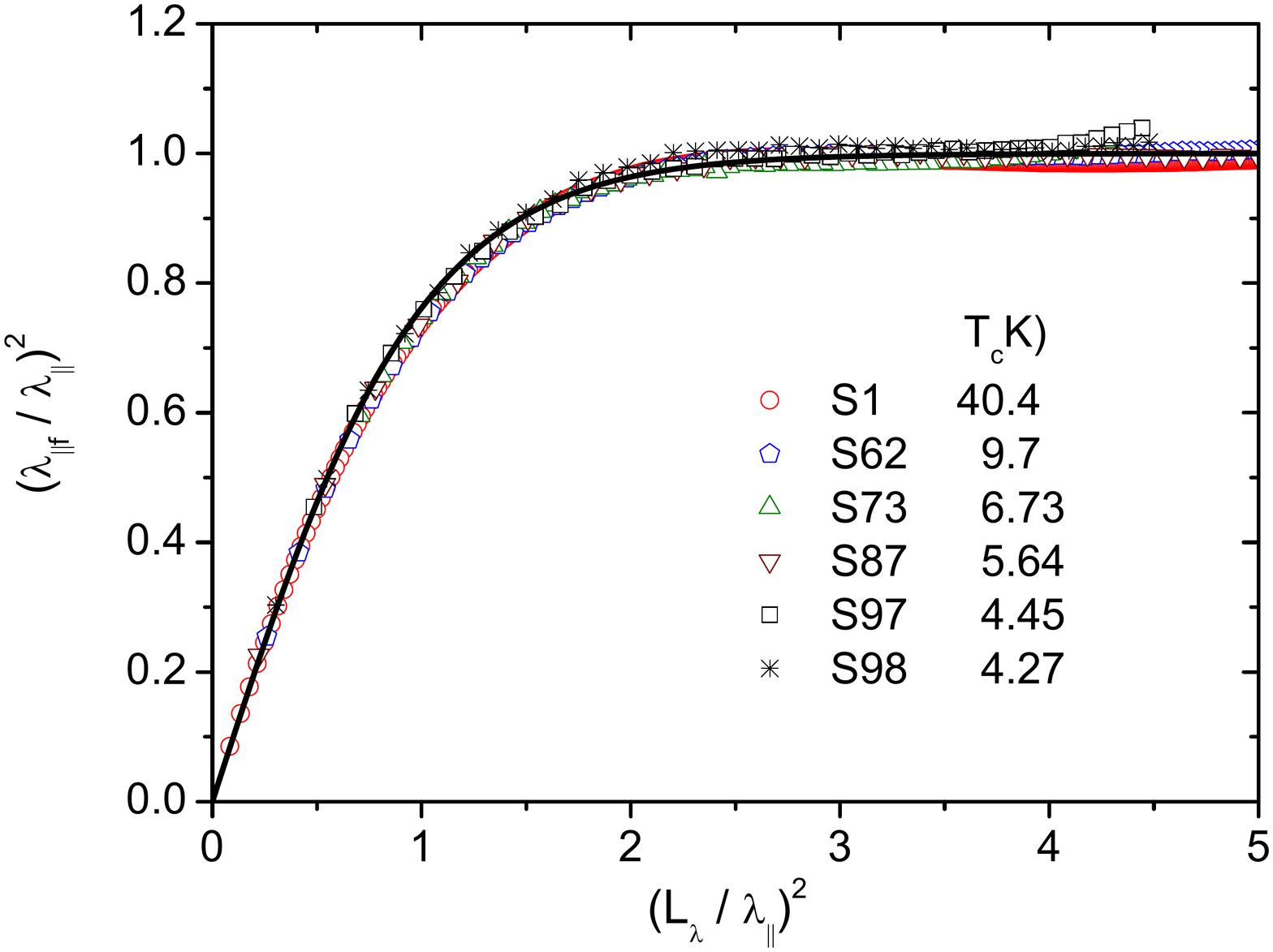}
\caption{Finite size scaling plot $\left( \lambda _{\parallel
f}\left( T\right) /\lambda _{\parallel }\left( T\right) \right) ^{2}$
\textit{vs }$\left( L_{\lambda }/\lambda _{\parallel }\left( T\right)
\right) ^{2}$ of the films listed in Table \ref{table1}. The line is $\left(
\lambda _{\parallel F}\left( T\right) /\lambda _{\parallel }\left( T\right)
\right) ^{2}=$tanh$\left( L_{\lambda }/\lambda _{\parallel }\left( T\right)
\right) ^{2}$.}
\label{fig5}
\end{figure}

 Finally we crosscheck the finite size scaling results. In Fig. \ref{fig5} we
depicted the finite size scaling plot $\lambda _{\parallel F}\left( T\right)
/\lambda _{\parallel }\left( T\right) )^{2}$ \textit{vs }$\left( L_{\lambda
}/\lambda _{\parallel }\left( T\right) \right) ^{2}$ of the films listed in
Table I. Given the estimates for $T_{c}$, $1/\lambda _{\parallel 0}^{2}$ and
$L_{\lambda \parallel }^{2}$ this plot should then let $\lambda _{\parallel
F}\left( T\right) /\lambda _{\parallel }\left( T\right) )^{2}$\ collapse
onto a single line. As Fig. \ref{fig5} shows we have a good collapse of the
data onto the scaling function $g\left( x\right) $ (Eq. (\ref{eq8})). For
this to happen, the estimates for $T_{c}$, $1/\lambda _{\parallel 0}^{2}$
and $L_{\lambda \parallel }^{2}=L_{c}\Lambda /T_{c}$ need to be correct. In
addition it uncovers the nature of the limiting length. If
$L_{c}=L_{ci}<L_{cf}$ it is traced back to inhomogeneities and if $%
L_{c}\simeq L_{cf}$ to the effective film thickness $L_{cf}$. Notably, this
collapse is almost perfect regardless of the nature of the limiting length.
In the films S1, S87, S97 and S98 it is the film thickness $L_{cf}$, while
in the films S62 and S73 inhomogeneities are what causes the limiting length
$L_{ci}<L_{cf}$ (see Table \ref{table1}). Nevertheless, the films attain the
critical regime where $g(x)\rightarrow 1$. The reason for this is that the
amplitude of the transverse correlation length $\xi _{\perp 0}^{t}$ remains
small compared to the film thickness $L_{cf}=6.6$~nm (Eq. (\ref{eq7b}), Table \ref{table1}).

\begin{figure}[t]
\includegraphics[width = 1 \columnwidth]{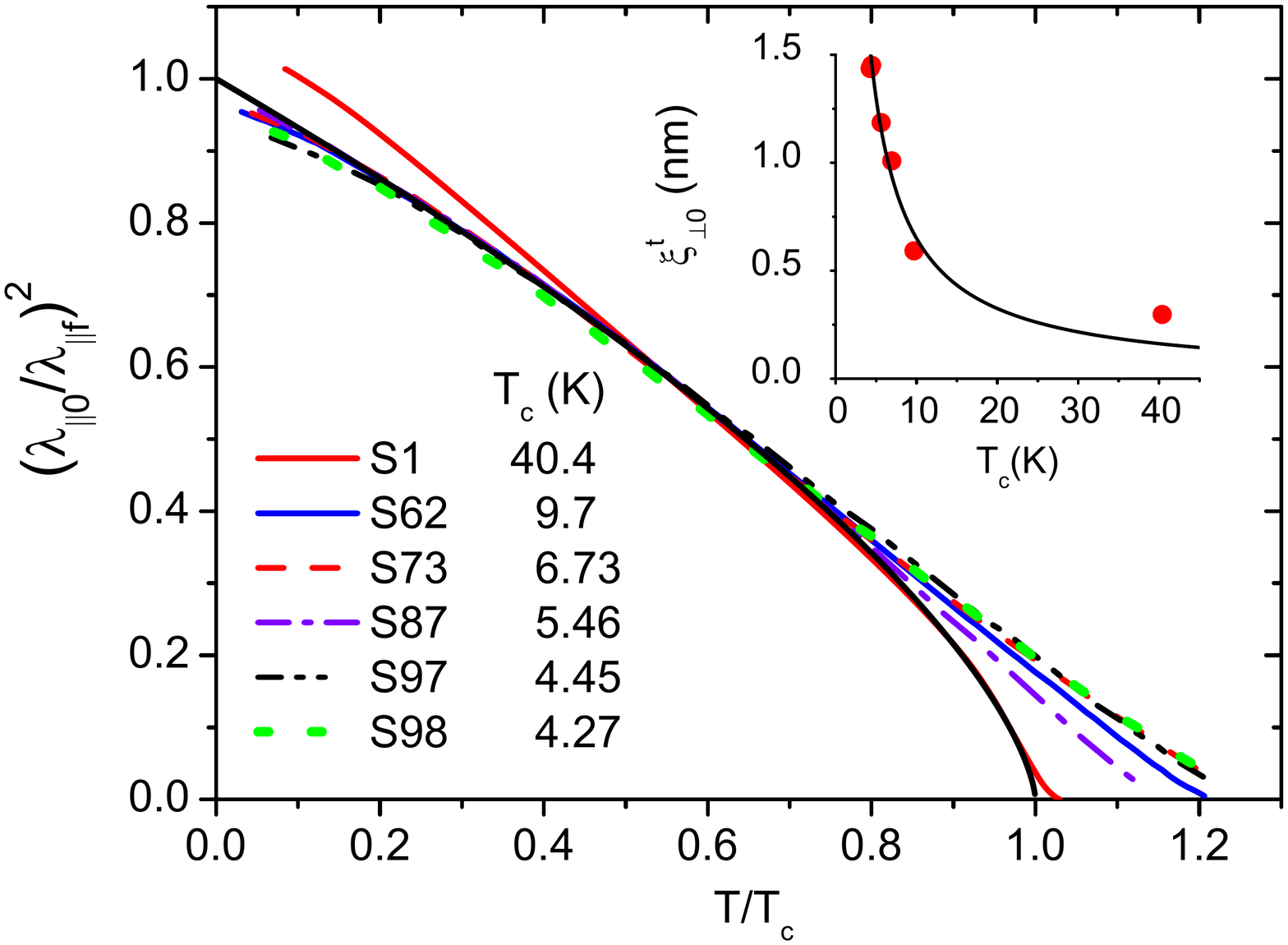}
\caption{Scaling plots $\left( \lambda _{\parallel 0}/\lambda _{\parallel
f}\left( T\right) \right) ^{2}$ \textit{vs }$T/T_{c}$ of the films listed in
Table \ref{table1} using the estimates for $T_{c}$ and $\lambda _{\parallel
0}$. The data collapsing on $\left( \lambda _{\parallel 0}/\lambda
_{\parallel }\left( T\right) \right) ^{2}=(1-T/T_{c})^{2/3}$ (black curve)
uncover the consistency with $3$D-xy criticality. The inset shows the $T_{c}$
dependence of the amplitude of the transverse correlation length $\xi
_{\perp 0}^{t}$ according to $\xi _{\perp }^{t}\left( t\right) =\xi _{t\perp
0}t^{-2/3}\propto t^{-2/3}/T_{c\text{ }}$(Eq. (\ref{eq20})). The line is $%
\xi _{\perp 0}^{t}=6.5/T_{c}$ with $\xi _{\perp 0}^{t}$ in nm and $T_{c}$ in K.}
\label{fig6}
\end{figure}

\begin{figure}[t]
\includegraphics[width = 1 \columnwidth]{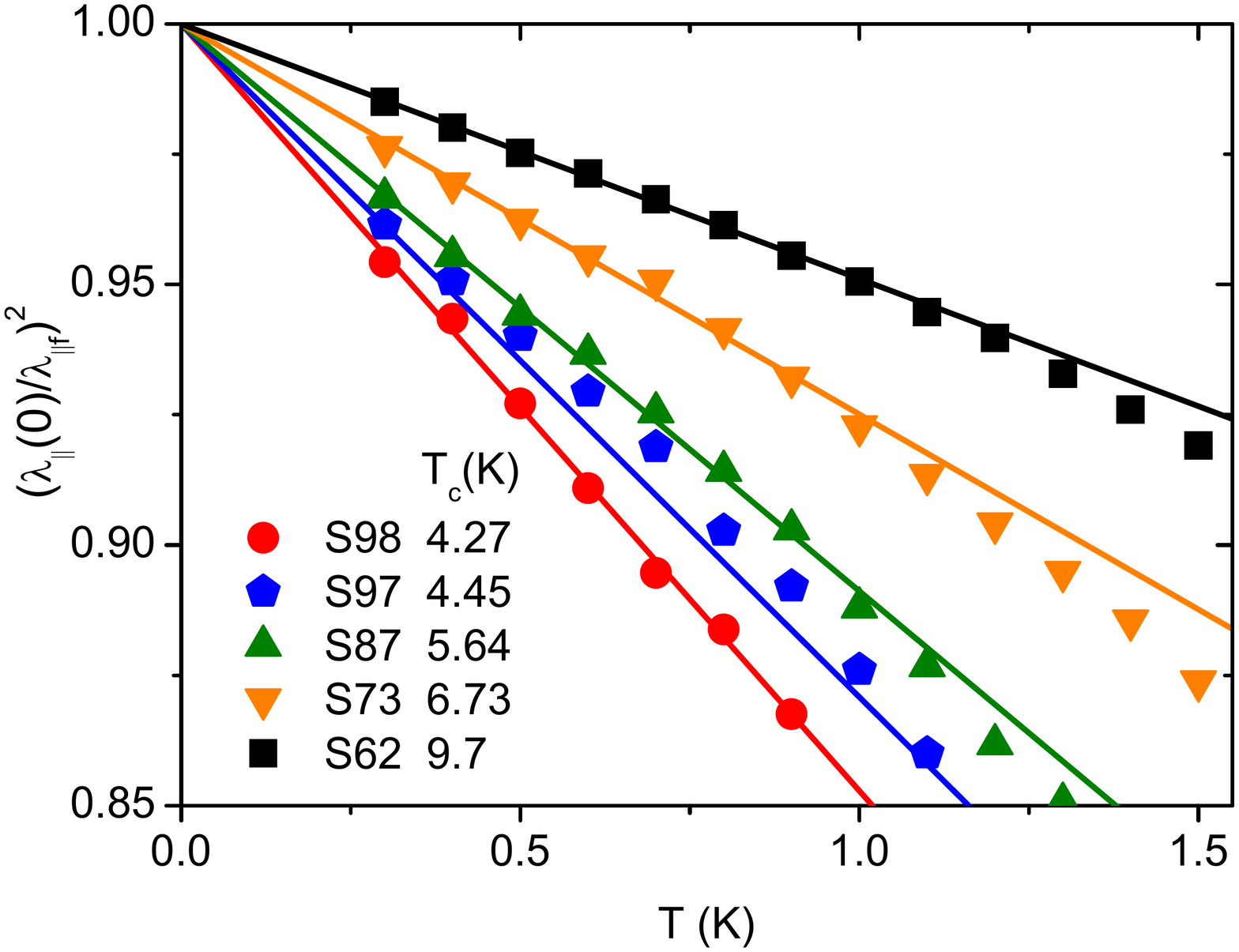}
\caption{$\lambda _{\parallel }^{2}\left( 0\right) /\lambda _{\parallel
f}^{2}\left( T\right) $ \textit{vs}. $T$ for the quoted films. The lines
mark the fits to $\lambda _{\parallel }^{2}\left( 0\right) /\lambda
_{\parallel }^{2}\left( T\right) =1-\alpha \lambda _{\parallel }^{2-}\left(
0\right) T$ \ yielding the estimates for $\lambda _{\parallel }^{2-}\left(
0\right) $ and $\alpha $ listed in Table \ref{table1} using the data from $0.3$ to $1$~K.}
\label{fig7}
\end{figure}

\begin{figure}[t]
\includegraphics[width = 1 \columnwidth]{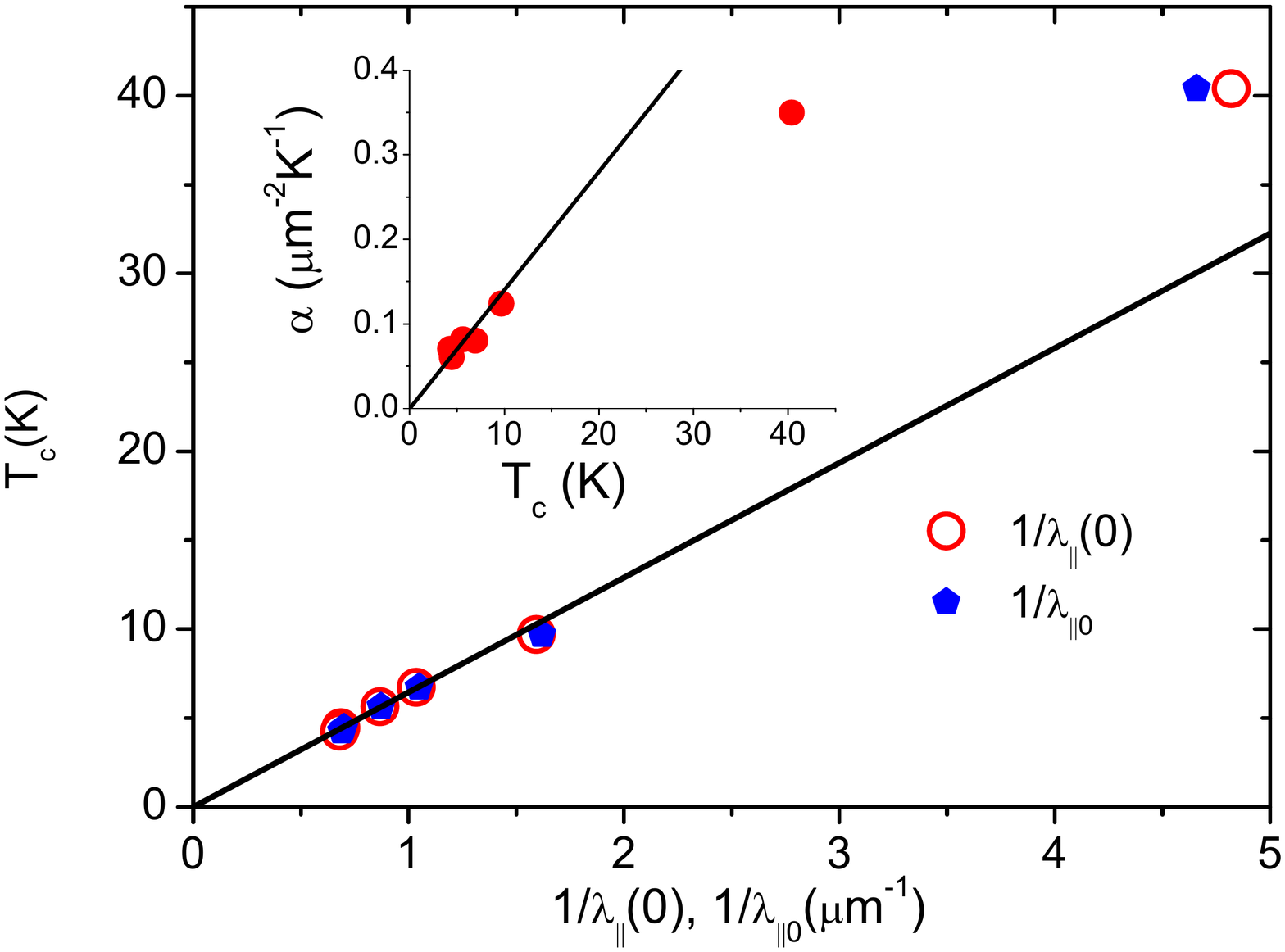}
\caption{$T_{c}$ \textit{vs}. $1/\lambda _{\parallel }\left( 0\right) $ and $%
1/\lambda _{\parallel 0}$ using the estimates listed in Table \ref{table1}. The solid
black line is $T_{c}=f_{\Vert }\left( 0\right) /\lambda _{\parallel }\left(
0\right) $ with\ $f_{\Vert }\left( 0\right) =6.45$~K$\mu $m. The inset shows $%
\alpha $ \textit{vs}. $T_{c}$. The straight line is $\alpha =\left(
y_{c}/f_{\Vert }^{2}\left( 0\right) \right) T_{c}$(Eq. (\ref{eq18})) with $%
y_{c}/f_{\Vert }^{2}\left( 0\right) =0.014\mu $m$^{-2}$~K$^{-1}$.}
\label{fig8}
\end{figure}

Next we turn to the analysis of the low-temperature penetration depth data
extending down to $0.3$~K. Therefore, an extrapolation is necessary to
estimate $\lambda _{\parallel }^{-2}\left( 0\right) $. Fig. \ref{fig7} shows
that the linear temperature dependence $\lambda _{\parallel }^{-2}\left(
T\right) =\lambda _{\parallel }^{-2}\left( 0\right) -\alpha T$, using the
data up to $1$~K, fits the data remarkably well. In $\lambda _{\parallel
}^{-}\left( T\right)$ it corresponds to $\left. d\lambda _{\parallel
}^{-}\left( T\right) /dT\right\vert _{T\rightarrow 0}=\alpha \lambda
_{\parallel }^{3}\left( 0\right) /2$ attributed by several authors to phase
fluctuations\cite{stroud,coffey}. In addition, the mapping on the (3 + 1)D-xy model implies a linear
temperature dependence because $\Upsilon _{\parallel }\left( T\right)
\propto \left\langle \left\vert \Psi \right\vert \right\rangle ^{2}\propto
\left( 1-T/T_{c}\right) $. The resulting $\lambda _{\parallel
}^{-2}\left( 0\right) $ and $\alpha $ are listed in Table \ref{table1}. With
the estimates for $T_{c},\lambda _{\parallel }^{-2}\left( 0\right)$,
$\lambda _{\parallel 0}^{-2}\left( 0\right) $ and $\alpha $ it is now
possible to clarify their compatibility with a quantum transition belonging
to the (3+1)D-xy universality class. Indeed, consistency with this
transition requires that the relationships $T_{c}=f_{\Vert }\left( 0\right)
/\lambda _{\parallel }\left( 0\right) =f_{\Vert }\left( 0\right) /\lambda
_{\parallel 0}$ (Eq. (\ref{eq19})), $\alpha =y_{c}T_{c}/f_{\Vert }^{2}\left(
0\right) $ (Eq. (\ref{eq18})), and $\lambda _{\parallel }^{2}\left( 0\right)
/\lambda _{\parallel }^{2}\left( T\right) =1-y_{c}T/T_{c}=1-\alpha \lambda
_{\parallel }^{2}\left( 0\right) T$ (Eq. (\ref{eq16})) are met. In Fig. \ref{fig8},
depicting $T_{c}$ \textit{vs}. $1/\lambda _{\parallel }\left(
0\right) $ and $1/\lambda _{\parallel 0}$, we observe that the first
relation is confirmed with
\begin{equation}
f_{_{\Vert }}\left( 0\right) \simeq 6.45~\text{K}\mu m\text{.}  \label{eq21}
\end{equation}%
Consequently, as $T_{c}$ drops quantum fluctuations reduce $1/\lambda
_{\parallel }\left( 0\right) $ and $1/\lambda _{\parallel 0}$ in the same
manner. Further consistency emerges from the $T_{c}$ dependence of the
coefficient $\alpha $ shown in the inset of Fig. \ref{fig8}, yielding with
$\alpha =\left( y_{c}/f_{\Vert }^{-2}\left( 0\right) \right) T_{c}$ the
estimates
\begin{equation}
y_{c}/f_{\Vert }^{2}\left( 0\right) \simeq 0.014~(\mu m^{-2}\text{K}^{-1})\text{,}
\label{eq22}
\end{equation}%
and with $f\left( 0\right) \simeq 6.45$~K$\mu m$%
\begin{equation}
y_{c}\simeq 0.58,  \label{eq23}
\end{equation}

for the universal coefficient. Noting that $\lambda _{\parallel }^{-2}\left(
T\right) =\lambda _{\parallel }^{-2}\left( 0\right) -\alpha T$ can be
rewritten in the form $\lambda _{\parallel }^{-2}\left( T\right) =\left(
T_{c}/f_{\Vert }\left( 0\right) \right) ^{2}-\alpha T_{c}\left(
T/T_{c}\right) $ it is evident that $\alpha $ scales in the limit
$T/T_{c}\rightarrow 0$ as $\alpha \propto T_{c}$.

\begin{figure}[t]
\includegraphics[width = 1 \columnwidth]{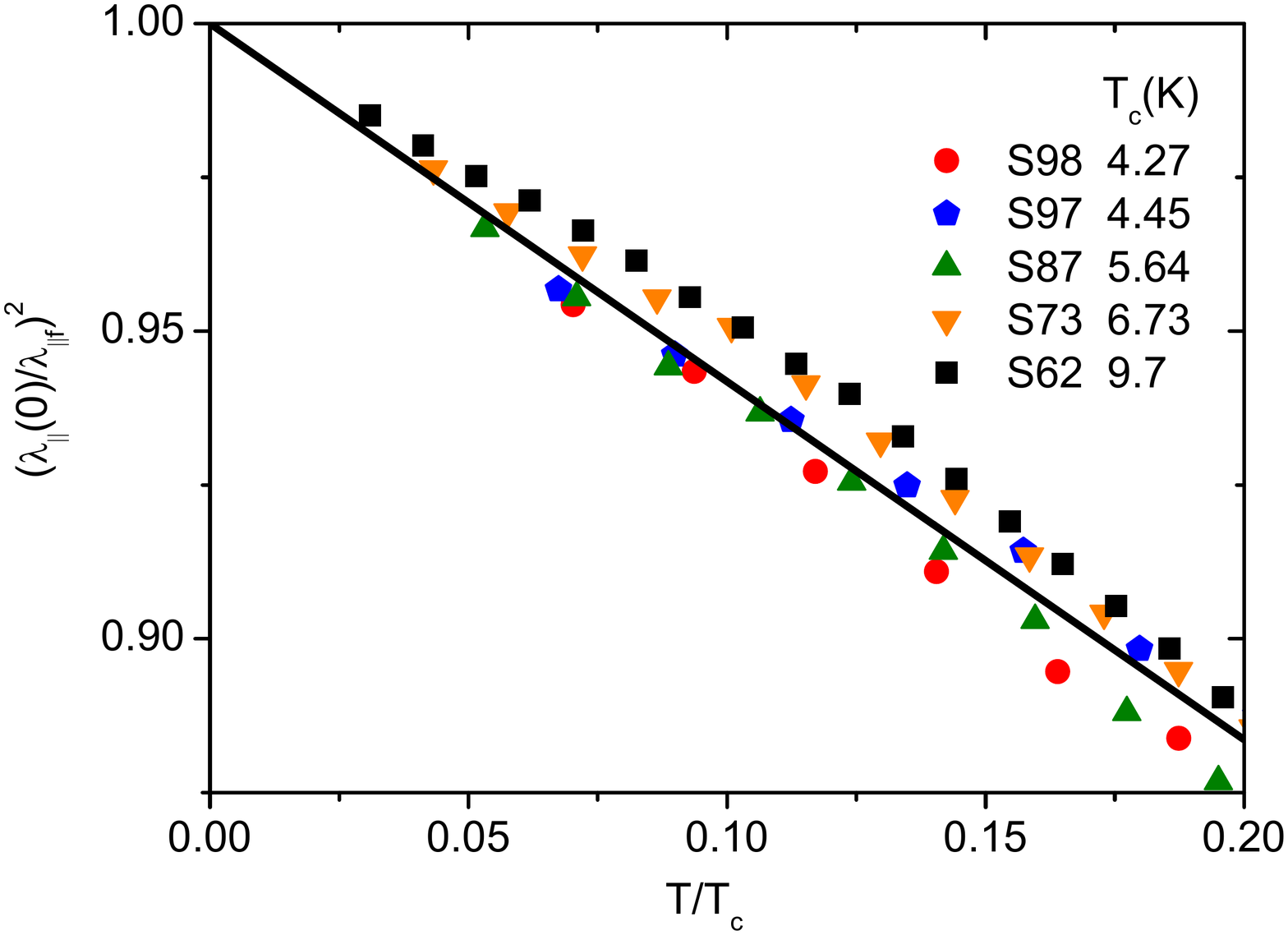}
\caption{Scaling plot $\lambda _{\parallel }^{2}\left( 0\right) /\lambda
_{\parallel f}^{2}\left( T\right) $ \textit{vs}. $T/T_{c}$ in the limit $%
T/T_{c}\rightarrow 0$. The line is $\lambda _{\parallel }^{2}\left( 0\right)
/\lambda _{\parallel }^{2}\left( T\right) =1-y_{c}T/T_{c}$ with the
universal coefficient $y_{c}=0.58$ (Eq. ( \ref{eq23})).}
\label{fig9}
\end{figure}

To complete the consistency checks we show in Fig. \ref{fig9} the plot
$\lambda _{\parallel }^{2}\left( 0\right) /\lambda _{\parallel f}^{2}\left(
T\right) $ \textit{vs}. $T/T_{c}$. The outlined linear temperature
dependence of $\lambda _{\parallel f}^{-2}\left( T\right) $ implies that the
data should fall on the line $\lambda _{\parallel }^{2}\left( 0\right)
/\lambda _{\parallel }^{2}\left( T\right) =1-y_{c}T/T_{c}$ with $y_{c}\simeq
0.58$ (Eq. (\ref{eq23})). The approach to this line is clearly seen as $%
T_{c}\rightarrow 0$. A crosscheck comes from the relation $T_{c}=f\left(
0\right) /\lambda _{\parallel }\left( 0\right) $ (Eq. (\ref{eq14})) after
which $T/T_{c}$ can be replaced\ by $T\lambda _{\parallel }\left( 0\right)
/f_{\Vert }\left( 0\right) $. This plot is shown in Fig. \ref{fig9}. Here
the data should collapse on the line $\lambda _{\parallel }^{2}\left(
0\right) /\lambda _{\parallel }^{2}\left( T\right) =1-uT\lambda _{\parallel
}\left( 0\right) $ with%
\begin{equation}
u=y_{c}/f_{\Vert }(0)=0.09,  \label{eq24}
\end{equation}
using the estimates given in Eqs. (\ref{eq21}) and (\ref{eq23}). Again, as
$T_{c}$ drops the data are seen to approach this line.

\begin{figure}[t]
\includegraphics[width = 1 \columnwidth]{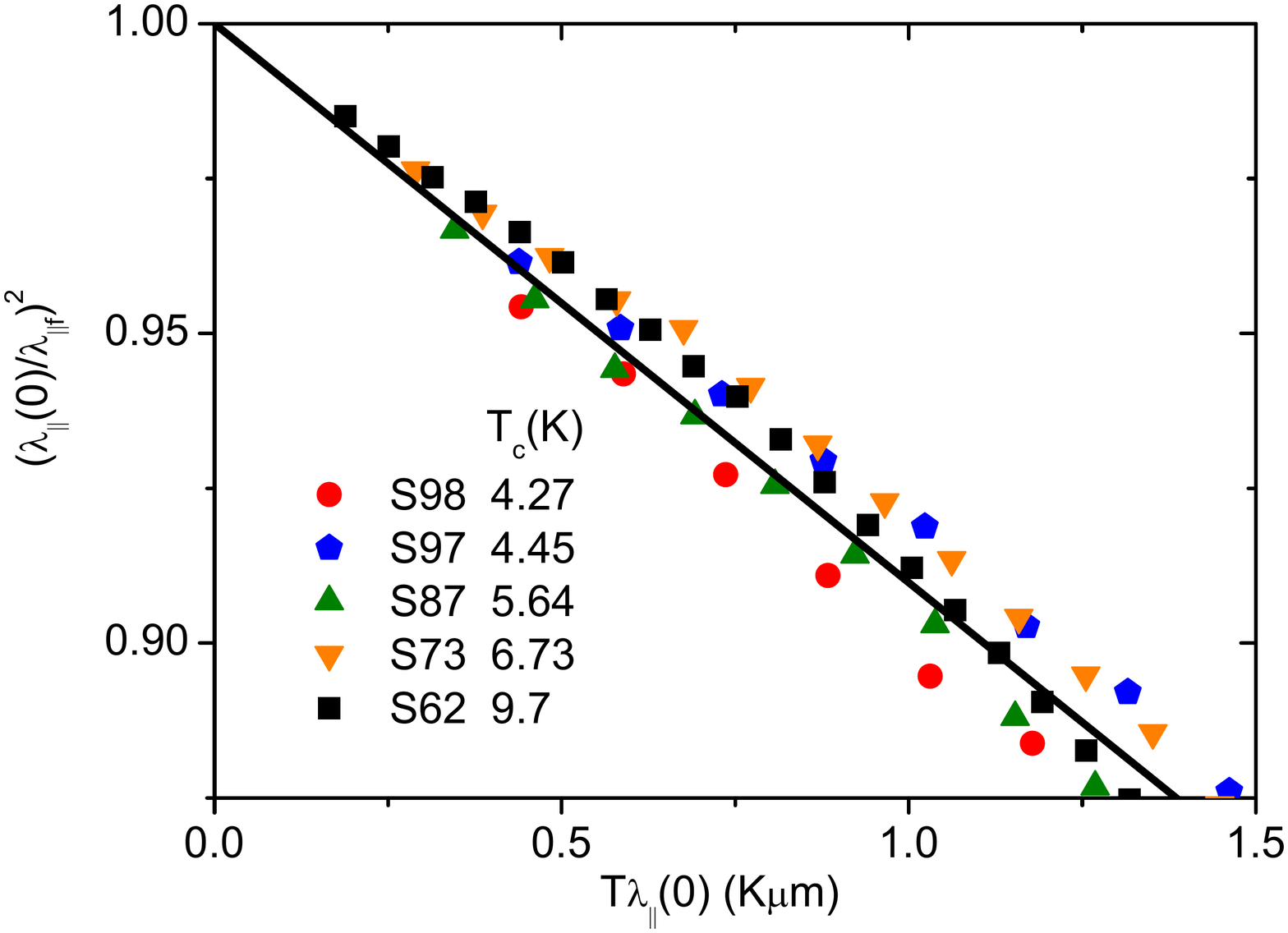}
\caption{Scaling plot $\lambda _{\parallel }^{2}\left( 0\right) /\lambda
_{\parallel f}^{2}\left( T\right) $ \textit{vs}. $T\lambda _{\parallel
}\left( 0\right) $ in the limit $T\lambda _{\parallel }\left( 0\right)
\rightarrow 0$. The line is $\lambda _{\parallel }^{2}\left( 0\right)
/\lambda _{\parallel }^{2}\left( T\right) =1-uT/T_{c}$ with $u=y_{c}/f\left(
0\right) \simeq 0.09$ (Eq. (\ref{eq24})).}
\label{fig10}
\end{figure}

\begin{figure}[t]
\includegraphics[width =1 \columnwidth]{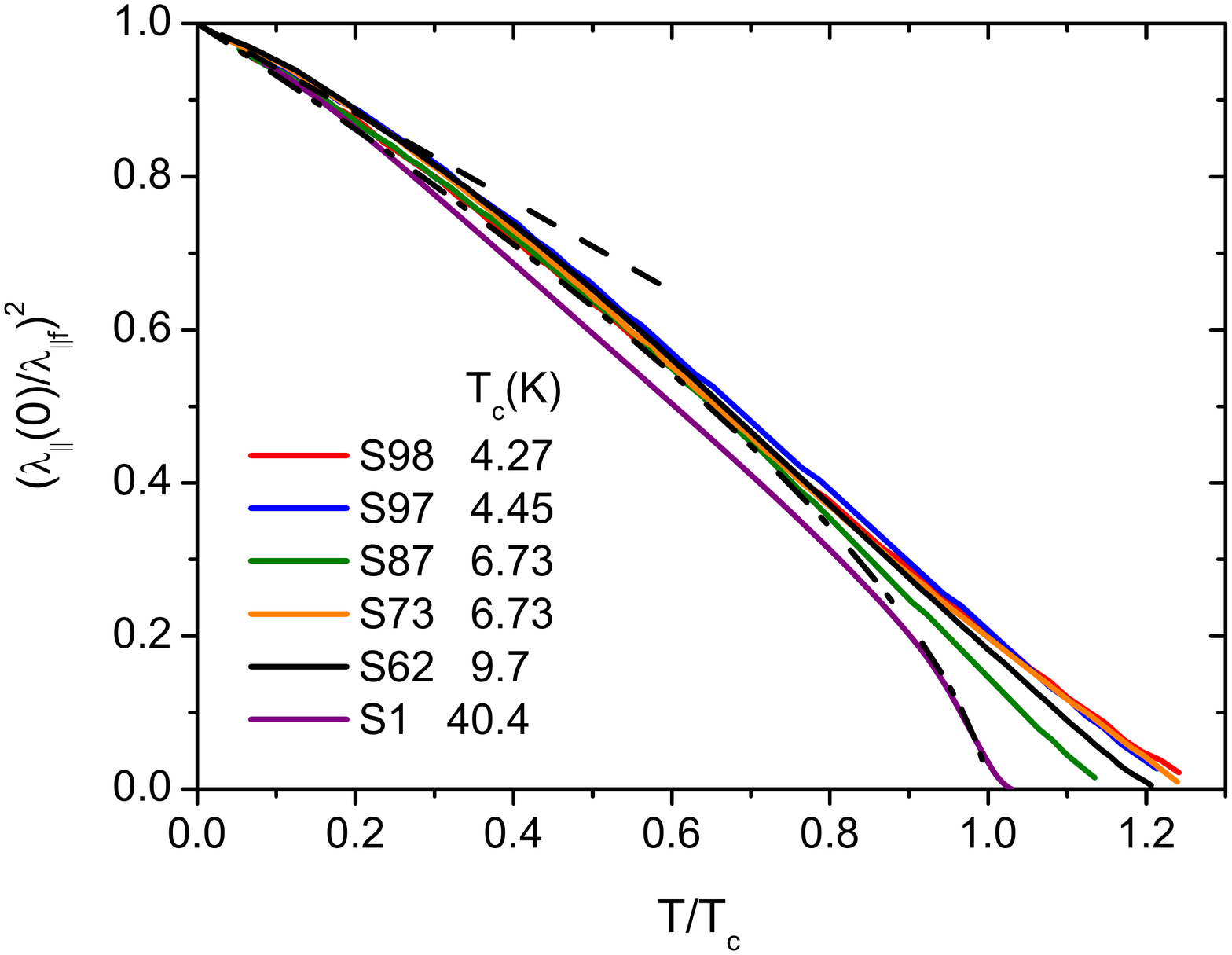}
\caption{Complete scaling plot $\lambda _{\parallel }^{2}\left( 0\right)
/\lambda _{\parallel f}^{2}\left( T\right) $ \textit{vs}. $T/T_{c}$. The the
dash- dot line marks the limiting quantum behavior $\lambda _{\parallel
}^{2}\left( 0\right) /\lambda _{\parallel }^{2}\left( T\right)
=1-y_{c}T/T_{c}$ with $y_{c}=0.58$ (Eq. (\ref{eq23})) and the dashed line $%
\lambda _{\parallel }^{2}\left( 0\right) /\lambda _{\parallel }^{2}\left(
T\right) =(1-T/T_{c})^{2/3}$ the thermal 3D-xy criticality with $\lambda
_{\parallel }\left( 0\right) =\lambda _{\parallel 0}$.}
\label{fig11}
\end{figure}

To complete the analysis of the selected overdoped La$_{2-x}$Sr$_{\mathbf{x}%
} $CuO$_{4}$ thin films we show in Fig. \ref{fig11} the plot $\lambda
_{\parallel }^{2}\left( 0\right) /\lambda _{\parallel f}^{2}\left( T\right) $
\textit{vs}. $T/T_{c}$. It contains the data from $0.3$~K up to the
temperature defined by the onset of the Meissner effect. Because the limit
$T/T_{c}\simeq 1$ has already been dealt with (see Fig. \ref{fig6}), we are
concentrating on the limit $T/T_{c}\rightarrow 0$. It can be reached along
the phase transition line $T_{c}\left( 1/\lambda _{\parallel }\left(
0\right) \right) $ where $\lambda _{\parallel }^{2}\left( 0\right) /\lambda
_{\parallel }^{2}\left( T\right) =t^{2/3}$ and $\lambda _{\parallel
0}\rightarrow \lambda _{\parallel }\left( 0\right) ,$ or along the line
$1/\lambda _{\parallel }\left( 0\right) \propto \delta ^{1/2}\propto T_{c}$
at $T=0$ where $\delta $ is the tuning parameter (Eq. (\ref{eq15a})). The
dashed line in Fig. \ref{fig11} marks this limit in terms of $\lambda
_{\parallel }^{2}\left( 0\right) /\lambda _{\parallel }^{2}\left( T\right)
=1-y_{c}T/T_{c}$ (Eq. (\ref{eq16})). Taking the Sr concentration $x$ in
La$_{2-x}$Sr$_{\mathbf{x}}$CuO$_{4}$ as tuning parameter it implies that the
empirical relation $T_{c}\propto \delta =x_{o}-x$ of Presland \textit{at al.}
\cite{presland} is violated as $x\rightarrow x_{o}$. Approaching the QPT
along the phase transition line $T_{c}\left( 1/\lambda _{\parallel }\left(
0\right) \right) $ the data is seen to collapse up to rather large $T/T_{c}$
values on the curve $\lambda _{\parallel }^{2}\left( 0\right) /\lambda
_{\parallel }^{2}\left( T\right) =t^{2/3}$ because $\lambda _{\parallel
0}$ $\rightarrow \lambda _{\parallel }\left( 0\right) $. However, as
$T/T_{c}\rightarrow 1$ the finite size induced rounding of the transition
sets in.

What remains is the question of which material parameters control the 3D-xy
transition. Fundamental parameters are the anisotropy $\gamma _{0}=\xi
_{\Vert 0}^{t}\left( 0\right) /\xi _{\perp 0}^{t}$ and the ratio $\kappa
_{\Vert 0}=\lambda _{\parallel 0}/\xi _{\Vert 0}^{t}$. Noting that the
universal relation (\ref{eq3}) transforms with these parameters to%
\begin{equation}
\gamma _{0}\kappa _{\Vert 0}=\Lambda /T_{c}\lambda _{\parallel 0}
\label{eq25}
\end{equation}
it is evident that $T_{c}\lambda _{\parallel 0}$ is fixed by $\gamma
_{0}\kappa _{\Vert 0}$. In the films considered here we obtain with
$f_{_{\Vert }}\left( 0\right) \simeq f_{_{\Vert 0}}\simeq 6.45$~K$\mu m$ (Eq. (\ref{eq21}))

\begin{equation}
\gamma _{0}\kappa _{\Vert 0}=\Lambda /T_{c}\lambda _{\parallel 0}=\Lambda
/f_{_{\Vert 0}}\simeq 91.5,  \label{eq26}
\end{equation}
and with $\gamma _{0}\simeq 10.6$ \cite{kohout}, $\kappa _{\Vert 0}\simeq
91.5 $ as $T_{c}\rightarrow 0$. This differs drastically from the behavior
of highly underdoped YBa$_{2}$Cu$_{3}$O$_{6+y}$, where $f_{_{\Vert
0}}\approx 6.5$~K$\mu m$ \cite{brown} but $\gamma _{0}\approx 100$ as
$T_{c}\rightarrow 0$ \cite{hosseini} and with that $\kappa _{\Vert 0}\approx 10
$ as $T_{c}\rightarrow 0$.

\section{Summary and Conclusions}

In summary, our analysis of the penetration depth data from Bo\v{z}ovi\'{c}
\textit{et al.}\cite{bosovic} for thin overdoped La$_{2-x}$Sr$_{\mathbf{x}}$%
CuO$_{4}$ films revealed: First, the observed mean-field-like behavior of
$\lambda _{\parallel f}^{-2}\left( T\right) $ is attributable to a finite
size effect. It stems from the fact that in an infinite and homogeneous
3D-xy system the transverse correlation length diverges as $\xi _{\perp
}^{t}\left( T\right) =\xi _{\perp 0}^{t}t^{-\nu }$, where $t=\left(
1-T/T_{c}\right) $ and $\nu \simeq 2/3$, while in the films $\xi _{\perp
f}^{t}\left( T_{c}\right) $ the divergence is cut off. In homogenous films
the limiting length is set by the effective film thickness $L_{cf}$ \ and in
the inhomogeneous ones by $L_{ci}<L_{cf}$, the size of the homogenous
domains. Nevertheless, in the films analyzed here, the $3$D-xy critical
regime is reached. The finite size scaling analysis of the rounded
transition provided estimates for the $T_{c}$'s and the critical amplitudes
$\lambda _{\parallel 0}\left( T_{c}\right) $ of the infinite and homogenous
counterpart. Because the $T_{c}$'s are lower than those estimated from the
the onset of the Meissner effect \cite{bosovic}, the mean-field-like
behavior of $\rho _{\Vert sf}\left( T\right) =\lambda _{\Vert f}^{-2}\left(
T\right) $ was traced back to a finite size effect. In addition the
estimates for $T_{c}$ and $\lambda _{\parallel 0}$ uncovered the
relationship $T_{c}=f_{\Vert 0}/\lambda _{\parallel 0}$. Second, having
established that the films exhibit 3D-xy critical behavior, rounded by the
finite size effect, we performed the analysis of the low temperature data
extending down to $0.3$~K. Therefore, an extrapolation was necessary to
estimate $\lambda _{\parallel }^{-2}\left( 0\right) $. It was shown that the
linear extrapolation with $\lambda _{\parallel f}^{-2}\left( T\right)
=\lambda _{\parallel }^{-2}\left( 0\right) -\alpha T$ fits the data up to $1$%
K very well. Given the estimates of $T_{c}$, $\lambda _{\parallel
}^{-2}\left( 0\right) ,$ $\alpha $ , and the $T_{c}$ dependence of the
latter, their consistency with a QPT mapping on the $(3+1)$D-xy
universality class was tested. This mapping requires that the
scaling relations $T_{c}=f_{\Vert }(0)/\lambda _{\parallel }\left( 0\right) $%
, $\alpha =y_{c}T_{c}/f_{\Vert }^{2}\left( 0\right) $ are satisfied and the
data plotted as $\lambda _{\parallel }^{2}\left( 0\right) /\lambda
_{\parallel f}^{2}\left( T\right) $ \textit{vs}. $T/T_{c}$ collapses on a
single curve. In the limit $T/T_{c}\rightarrow 0$ and sufficiently small
$T_{c}$ it should tend to the line $\lambda _{\parallel }^{2}\left( 0\right)
/\lambda _{\parallel f}^{2}\left( T\right) =1-y_{c}T/T_{c}$ and in the
$T/T_{c}\rightarrow 1$ limit to $t^{2/3}$. However, due to the rounded
transition is this limit not fully attainable and $\lambda _{\parallel
}^{2}\left( 0\right) /\lambda _{\parallel f}^{2}\left( T\right) $ exceeds
$T/T_{c}=1$. Nevertheless, as $T_{c}$ drops the collapse improved because
$\lambda _{\parallel }^{-2}\left( T\right) =\lambda _{\parallel 0}^{-2}t^{2/3}
$ approached $\lambda _{\parallel }^{-2}\left( T\right) =\lambda _{\parallel
}^{-2}\left( 0\right) t^{2/3}$, due to the crossover from the thermal to the
quantum fluctuation dominated regime. The consistency shown with these
characteristics of 3D-xy- and (3+1)D-xy criticality clearly revealed that
the suppression of the superfluid density in the thin and overdoped La$_{2-x}
$Sr$_{\mathbf{x}}$CuO$_{4}$ films of Bo\v{z}ovi\'{c} \textit{et al.} \cite%
{bosovic} is a quantum effect. It uncovers the crossover from the thermal to
the quantum fluctuation dominated regime as $T_{c}$ drops. In the
classical-quantum mapping it is the crossover from $3$D to $(3+1)$D. The
thermal regime shows finite size limited 3D-xy criticality and the quantum
regime is compatible with ($3+1$)D-xy criticality.
  Another manifestation of the relevance of quantum fluctuations comes
according to Eq. (\ref{eq20b}) from the observation that in the overdoped La$%
_{2-x}$Sr$_{\mathbf{x}}$CuO$_{4}$ films conductivity measurements show that
below $T_{c}$ a large fraction of the Drude weight remains uncondensed \cite
{mahmood}. Moreover, consistent with the importance of thermal fluctuations
(Eq. (\ref{eq20a})) in the overdoped regime are Nernst, torque, magnetization
\cite{Lu}, recent specific heat and ARPES measurements \cite{He}, revealing
in overdoped Bi-2212 single crystals Cooper pairs which are already formed
above $T_{c}$. However, it should be kept in mind that the observed strong
diamagnetism and Nernst signal in a wide temperature window above $T_{c}$
exceeds the fluctuation dominated regime drastically. Indeed, taking the
inhomogeneities of single crystals into account, the effective
dimensionality is reduced to zero and reduced dimensionality enhances the
thermal fluctuations \cite{wey}. Nonetheless, the proven existence of copper pairs above $T_{c}$ implies that
they contribute to the transport properties and with that modify the
standard Fermi liquid picture.

\begin{acknowledgments}
I thank Stefano Gariglio and Jean-Marc Triscone for their support.
\end{acknowledgments}

\bibliography{ref}

\end{document}